\documentclass[journal=jctcce,amsmath,amssymb,manuscript=article]{achemso}
\usepackage{graphicx}
\usepackage{dcolumn}
\usepackage{multirow}

\usepackage{amsmath}
\usepackage{subfigure}
\usepackage{braket}
\usepackage{color}
\usepackage{epstopdf}
\usepackage[normalem]{ulem}
\usepackage{ulem}
\usepackage{listings}
\usepackage{import}
\usepackage{xcolor}
\usepackage{framed}
\usepackage{mdwlist}
\usepackage{diagbox}
\usepackage{eufrak}
\usepackage{bm}

\usepackage{algorithm}
\usepackage{algorithmicx}
\usepackage{algpseudocode}

\newcommand{\defeq}{\ensuremath{\stackrel{\text{def}}{=}}}
\newcommand{\tr}{\text{tr}}
\newcommand{\pair}[2]{\ensuremath{\braket{#1,#2}}}
\newcommand{\Neighbour}[1]{\mathfrak{N}[#1]}

\newcommand{\suml}[0]{\sum\limits}
\newcommand{\bigcupl}[0]{\bigcup\limits}
\newcommand{\argmin}[0]{\text{argmin}}

\renewcommand{\H}[0]{\ensuremath{\Sigma^{\textnormal{X}}}}
\newcommand{\Ht}[0]{\ensuremath{\mathcal{H}^{\textnormal{X}}}}
\newcommand{\R}[0]{\mathbf{R}}

\renewcommand{\r}[0]{\mathbf{r}}
\newcommand{\0}[0]{\mathbf{0}}
\renewcommand{\k}[0]{\mathbf{k}}
\newcommand{\aR}[1]{\tilde{#1}}

\newcommand{\Patom}[1]{\underline{#1}}

\newcommand{\bfr}{ {\bf r}} 
\newcommand{\bfR}{ {\bf R}} 
\newcommand{\bfk}{ {\bf k}} 

\newcommand{{\VASP}}{\textsc{vasp}}
\newcommand{{\FHIaims}}{\textsc{fhi-aims}}
\newcommand{{\ABACUS}}{\textsc{abacus}}
\newcommand{{\QEcode}}{\textsc{qe}}

\author{Peize Lin}
\affiliation{CAS Key Laboratory of Quantum Information, University of Science and Technology of China, Hefei, Anhui 230026, China}
\author{Xinguo Ren}
\email{renxg@iphy.ac.cn}
\affiliation{Institute of Physics, Chinese Academy of Sciences, Beijing 100190, China}
\author{Lixin He}
\email{helx@ustc.edu.cn}
\affiliation{CAS Key Laboratory of Quantum Information, University of Science and Technology of China, Hefei, Anhui 230026, China}

\title{Efficient Hybrid Density Functional Calculations for Large Periodic Systems Using Numerical Atomic Orbitals}

\abbreviations{IR,NMR,UV}

\begin{document}

\begin{tocentry}
%
%
%
%
\centering
\includegraphics[width=0.6\textwidth]{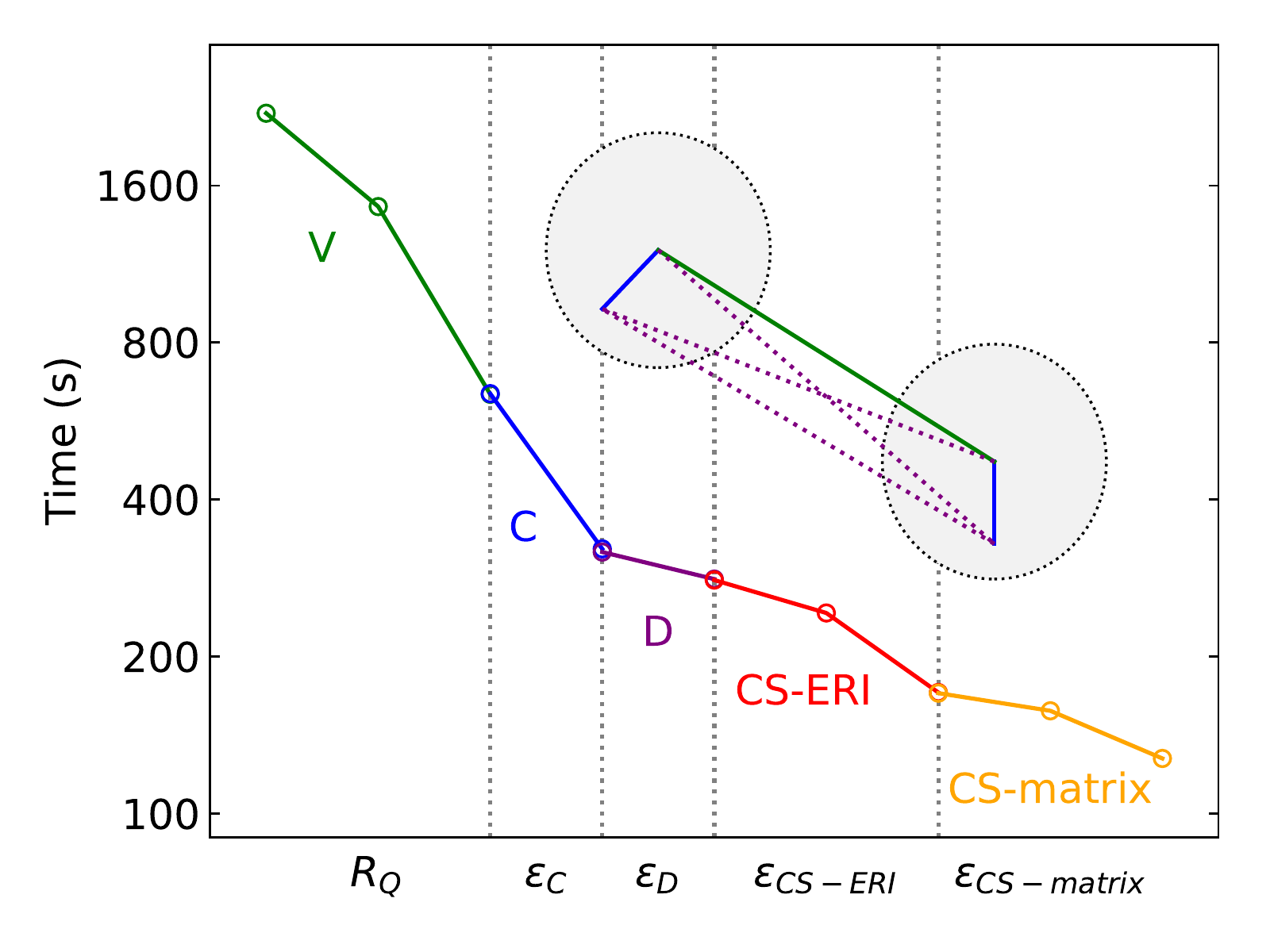}

\end{tocentry}

\begin{abstract}
We present an efficient, linear-scaling implementation for building the (screened) Hartree-Fock exchange (HFX) matrix
for periodic systems within the framework of numerical atomic orbital (NAO) basis functions.
Our implementation is based on the localized resolution of the identity approximation
by which two-electron Coulomb repulsion integrals can be obtained by only computing two-center quantities
-- a feature that is highly beneficial to NAOs.
By exploiting the locality of basis functions and efficient prescreening of the intermediate three- and two-index tensors,
one can achieve a linear scaling of the
computational cost for building the HFX matrix with respect to the system size.
Our implementation is massively parallel,
thanks to a  MPI/OpenMP hybrid parallelization strategy for distributing the computational load and memory storage.
All these factors add together to enable highly efficient hybrid functional calculations for large-scale periodic systems.
In this work we describe the key algorithms and implementation details for the HFX build as implemented in the \textsc{ABACUS} 
code package. The performance and scalability of our implementation
with respect to the system size and the number of CPU cores are demonstrated for selected benchmark systems up to 4096 atoms.

\end{abstract}




\section{Introduction}
Hybrid density functionals (HDFs) \cite{Becke:1993} belong to the fourth rung of the Jacob's ladder \cite{Perdew/Schmidt:2001}
in Kohn-Sham (KS) density functional theory (DFT) \cite{Hohenberg/Kohn:1964,Kohn/Sham:1965}.
On the one hand,
compared to lower-rung local- and semi-local approximations,
HDFs, as implemented in the framework of generalized KS (gKS) theory \cite{Seidl/etal:1996},
typically deliver better accuracy for ground-state properties and,
additionally, provide physically more sound single-particle energy spectra.
On the other hand, compared to the correlated methods (e.g., fifth-rung functionals,
the quantum chemistry methods, and Green-function based many-body approaches),
HDFs also have clear advantages.
This is not only because HDFs are significantly cheaper than the correlated methods,
but also because they are able to offer a range of useful properties in a single calculation,
including e.g., electron density, ground-state energies, atomic structures,
and single-particle orbital energies, and so on.
Not all of these properties are easily accessible from the more expensive many-body approaches.
As such, HDFs have been widely used in quantum chemistry
\cite{Becke:1993,Perdew/Ernzerhof/Burke:1996,Ernzerhof/Scuseria:1999,Adamo/Barone:1999,Zhao/Truhlar_M06:2008,Chai/Head-Gordon:2008},
and are becoming increasingly popular in computational materials science \cite{Heyd/Scuseria/Ernzerhof:2003}.

An essential feature of HDFs is that a portion of Hartree-Fock-type
exchange (HFX) is incorporated in the
construction of the exchange-correlation (XC) functional.
Furthermore, in recently developed HDFs,
the range-separation framework \cite{Toulouse/Colonna/Savin:2004} is often invoked,
in which the HFX is decomposed into a short-range part and a long-range part,
and different portions of the two parts are utilized in the functional constructions
\cite{Likura/etal:2001,Heyd/Scuseria/Ernzerhof:2003,Yanai/Tew/Handy:2004,Vydrov/Scuseria:2006,Chai/Head-Gordon:2008,Skone/etal:2016}.
Another interesting development is that
the initially empirical parameters in HDFs are made system-dependent and determined on the fly
\cite{Shimazaki/Asai:2008,Shimazaki/Asai:2010,Marques/etal:2011,Shimazaki/Nakajima:2014,Skone/etal:2014,Skone/etal:2016,He/Franchini:2017},
or satisfying certain physical constrains \cite{Kronik/etal:2012,Atalla/etal:2013}.
In particular, the concept of double screening mechanism involving both the dielectric
screening and metallic screening channel is instructive for developing more refined next-generation
DHFs \cite{Shimazaki/Asai:2008,Shimazaki/Asai:2010,Shimazaki/Nakajima:2014,Cui/etal:2018}.
All these efforts have increased the flexibility and application territories of HDFs.
For a recent review, see Ref.~\cite{Zhang/Jiang_etal:2020}.

Computationally, the build of the full-range or range-separated HFX matrix,
constitutes the major bottleneck for HDF calculations.
The canonical scaling of the computational cost
of this step is ${\cal O}(N^4)$ with $N$ being the system size.
Historically, efficient algorithms for computing the HFX matrix were
first developed in quantum chemistry using atom-centered Gaussian-type
orbitals (GTOs).
Back then, ${\cal O}(N^2)$ scaling
\cite{Almlof/Faegri/Korsell:1982,Haser/Ahlrichs:1989}
and even linear-scaling algorithms
\cite{Burant/Scuseria/Frisch:1996,Schwegler/Challacombe:1996,Schwegler/Challacombe/Head-Gordon:1997,Ochsenfeld/White/Head-Gordon:1998,Rudberg/Rubensson/Salek:2011}
for building the HFX matrix have been designed,
by exploiting the locality of GTO basis functions,
and (for insulating systems) the sparsity of the density matrix.
To extend the application of HDFs to condensed matter systems,
it is customary to impose the periodic boundary condition (PBC) in the implementation.
In practice, this means one needs to treat a rather large supercell,
and handle additional complexities such as the translation symmetry
with respect to the lattice vectors and the singularity of the Coulomb operator.
Since the early periodic Hartree-Fock (HF)
implementation \cite{Dovesi/etal:1980,Pisani/etal:1988}
in the C\textsc{rystal} code \cite{Dovesi/etal:2014},
other GTO-based HF and HDF implementations have been reported \cite{guidon2008ab,Guidon/etal:2009,Paier/etal:2009},
with linear-scaling cost for building the HFX matrix.
As the interest in HDFs grows in the solid-state community,
more HFD implementations have been reported,
within the projector augmented wave (PAW) \cite{Paier/etal:2006}, pseudopotential plane-wave (PW),
and linearized augmented plane wave (LAPW) \cite{Betzinger/etal:2010} frameworks,
where the PBC is automatically satisfied.
However, due to the extended nature of the PW and LAPW basis functions,
the computational cost of the HFX in these implementations
follows a canonical ${\cal O}(N^4)$ scaling as the system size.
By a transformation from the Bloch orbitals to a localized Wannier function representation,
numerical techniques can again be developed to achieve a linear-scaling numerical cost of
the HFX \cite{Wu/Selloni/Car:2009} matrix construction.
Other techniques to speed up the HDF calculations have also been developed in the PW basis context,
including, e.g.,
the recently proposed adaptively compressed exchange operator technique \cite{LinLin:2016}.

In recent years, owing to their strict spatial locality and more realistic radial behavior,
the numerical atomic orbitals (NAOs) are gaining increasing popularity as the basis set choice in first-principles electronic structure calculations.

For ground-state calculations based on conventional local and semilocal approximations
\cite{Kohn/Sham:1965,Becke:1988a,Lee/Yang/Parr:1992,Perdew/Burke/Ernzerhof:1996},
the reliability and efficiency of the NAO basis set have been well established,
as been demonstrated by the flourishing of NAO-based first-principles computer code packages
\cite{Delley:2000,Koepernik/Eschrig:1999,Soler/etal:2002,Ozaki/etal:2008,Blum/etal:2009,Li/Liu/etal:2016}.

In case of HDFs,
where the computation of two-electron Coulomb repulsion integrals (ERIs) is a necessity,
NAO-based implementations have only appeared recently.
Existing implementations circumvent a
straightforward evaluation of ERIs in terms of NAOs in one way or another,
by expanding the NAOs in terms of GTOs \cite{Shang/Li/Yang:2010,Qin/etal:2015},
or by employing the resolution-of-the-identity (RI) technique \cite{Ren/etal:2012,Ihrig/etal:2015,Levchenko/etal:2015,Lin/Ren/He:2020}
(as well as its refined variant -- the interpolative separable density fitting scheme \cite{Lu/Yin:2015,Qin/Hu/Yang:2020}).
Within the RI
\cite{Feyereisen/Fitzgerald/Komornicki:1993,Vahtras/Almlof/Feyereisen:1993,Weigend/Haser/Patzelt/Ahlrichs:1998} -- also known as
variational density fitting \cite{Whitten:1973,Dunlap/Connolly/Sabin:1979,Dunlap/Rosch/Trickey:2010} -- approximation,
the four-index ERIs are decomposed into three- and two-index ones,
whereby the storage requirement and the computational cost are significantly reduced.
For molecular systems,
the accuracy and efficiency of the
conventional RI approximation based on the Coulomb metric
have been well established for the HFX and explicit correlation calculations using both GTOs
\cite{Weigend:2002,Weigend/Haser/Patzelt/Ahlrichs:1998,Eshuis/Yarkony/Furche:2010,DelBen/etal:2013} and NAOs \cite{Ren/etal:2012}.

Despite its nice properties such as the preservation of positive definiteness of the ERI matrix and the vanishing of
the expansion errors up to linear order \cite{Dunlap/Connolly/Sabin:1979, Merlot/etal:2013},
the conventional Coulomb-metric RI scheme is not particularly suitable for large molecules and extended systems.
For these systems, the conventional RI scheme,
which requires the computation of a very large number of three-center Coulomb integrals
and an inversion of a large Coulomb matrix,
can become prohibitively expensive.
To deal with this problem, localized variants of the RI approach have been developed
\cite{Billingsley/Bloor:1971,Pisani/etal:2005,Pisani/etal:2008,Sodt/etal:2006,Sodt/Head-Gordon:2008,Reine/etal:2008,Merlot/etal:2013,Ihrig/etal:2015},
where the products of two atomic orbitals (AOs) are expanded in terms of a limited set of auxiliary basis functions (ABFs)
centering in the neighborhood of the two AO centers.
Restricting the ABFs expanding the AO pair products to only the two center where the two AOs are located,
one obtains what we called the localized RI (LRI) scheme
\cite{Ihrig/etal:2015,Levchenko/etal:2015,Lin/Ren/He:2020},
or the so-called pair-atomic RI (PARI) approximation \cite{Merlot/etal:2013,Wirz/etal:2017} in the quantum chemistry literature.
The LRI scheme is essential for enabling efficient NAO-based periodic HDF implementations
\cite{Levchenko/etal:2015,Lin/Ren/He:2020}.

In a recent paper \cite{Lin/Ren/He:2020},
we benchmarked the accuracy of LRI for HDF calculations within a pseudopotential-based NAO framework.
Using the Heyd-Scuseria-Ernzerhof (HSE) screened HDF \cite{Heyd/Scuseria/Ernzerhof:2003} as implemented
in the \textsc{ABACUS} (Atomic-orbital Based Ab-initio Computation at UStc) code 
\cite{abacusweb,Chen/Guo/He:2010,Li/Liu/etal:2016},
we showed that, when paired with suitably constructed auxiliary basis set,
the errors in the computed band gap values incurred by LRI is below 10 meV for prototypical semiconductors and insulators.
Such an error is significantly smaller than the errors stemming from other sources,
such as the incompleteness of one-electron basis set and the quality pseudopotentials.
These tests indicate that, within the pseudopotential-based NAO framework,
the LRI approximation is able to provide adequate accuracy for practical purposes.

In this paper, we describe the details of the HDF implementation within the \textsc{ABACUS} code package 
\cite{abacusweb,Chen/Guo/He:2010,Li/Liu/etal:2016}.
\textsc{ABACUS} is a recently developed first-principles software that employs NAOs as basis functions and the
norm-conserving pseudopotentials \cite{hamann1979norm} to describe the ion cores.
An economic, linear-scaling implementation for building the HFX for large periodic systems has been carried out,
utilizing several numerical algorithms including the LRI scheme to compute the ERIs,
an efficient prescreening of the intermediate quantities to form the HFX matrix,
an elegant ``back-folding" procedure to account for the translational symmetry,
and a ``memorization" technique for efficiently handling the two-center Coulomb integrals.
Furthermore, depending on the actual size of the system to be simulated and the available resources,
we use either Machine-scheduling or K-means algorithm to distribute the computational load,
resulting in a parallel program showing excellent efficiency up to thousands of CPU cores.

The outline of this paper is as follows:
Section II briefly reviews the exact-exchange theory and the resolution of the identity scheme.
Section III describes all algorithms and techniques we have proposed to decrease the computing time and memory consumes.
Section IV shows the benchmark results about the accuracy and performance of the implementations.
Section V concludes this paper.

\section{Methods \label{sec:methods}}
In the present work,
the HFX matrix is evaluated within the framework of linear combination of atomic orbitals (LCAO).
Here, we explicitly consider periodic systems,
and an AO $\phi_i(\r)$ centering at an atom $\aR{I}$ within a unit cell $\R_I$ is denoted as
\begin{equation}
	\phi_{\aR{I}(\R_I)i}(\r) \defeq \phi_i(\r-{\bm \tau}_{\aR{I}}-\R_I)
	\label{eq:AO}
\end{equation}
where ${\bm \tau}_{\aR{I}}$ is the position of the atom $\aR{I}$ within the unit cell.
To simplify the notation, we omit the lattice vector index $\R_I$ in the following,
and denote that $\phi_{\aR{I}(\R_I)i}(\r)=\phi_{Ii}(\r)$, but keep in mind that different
atoms might be located in different unit cells.
In the present paper, we choose to adopt a notational system that is slightly different from the one
used in Ref.~\citenum{Lin/Ren/He:2020}, by which the indices of the atoms are clearly indicated.
As will become clear later, by specifying the individual atoms from which the basis functions originate,
our algorithm of evaluating the HFX matrix can be better explained.

Within the LCAO formulation, the HFX matrix for a periodic system is formally given by
\begin{equation}
	 \H_{Ii,Jj}
	= \sum_{K,L} \sum_{k\in K, l\in L}
	(\phi_{Ii} \phi_{Kk} | \phi_{Jj} \phi_{Ll}) D_{Kk,Ll}
\label{eq:HFX_ERIs}
\end{equation}
where we use the abbreviation
\begin{equation}
	(f|g) \defeq \iint f(\r) v(\r-\r') g(\r') d\r d\r'
	\label{eq:coulomb_integrals}
\end{equation}
to denote the interaction integral between two functions $f(\r)$ and $g(\r)$ via a potential
$v(\r-\r')$ and
$D_{Kk,Ll}$ is the density matrix in real space,
\begin{equation}
    D_{Kk,Ll}= \frac{1}{N_\k}\sum_{n,\k \in 1\textnormal{BZ}} f_{n\k} c_{k,n}(\k)c_{l,n}^\ast(\k)e^{-i\k \cdot (\R_K-\R_L)} \,.
    \label{eq:DM_realspace}
\end{equation}
In eq~\eqref{eq:DM_realspace}, $c_{k,n}(\k)$ are KS eigenvectors,
$f_{n\k}$ are the occupation numbers,
and $N_\k$ is the number of $\k$ points in the 1st Brillouin zone (1BZ).
Furthermore, the lattice vectors $\R_K$ and $\R_L$ label the unit cells
where the atom $K$ and $L$ are located, respectively.
Depending on the choice of the potential $v(\r-\r')$ in eq~\eqref{eq:coulomb_integrals},
one can either obtain the full-range HFX (when $v(\r-\r')=1/|\r-\r'|$)
or the short-range HFX (e.g., $v(\r-\r')=\textnormal{erfc}(\mu|\r-\r'|)/|\r-\r'|$,
with erfc$(x)$ being the complementary error function, and $\mu$ a range-separation parameter).
Since our algorithm presented below does not depend on the specific form of $v(\r-\r')$,
we don't distinguish full-range or short-range HFX, unless this turns out to be necessary.

In this work, we use the LRI approximation
\cite{Ihrig/etal:2015,Levchenko/etal:2015,Lin/Ren/He:2020},
also known as PARI \cite{Merlot/etal:2013,Wirz/etal:2017},
to evaluate the ERIs.
Within LRI, the product of two NAOs centering on two atoms is approximately expanded as
\begin{align}
	 \phi_{Ii}(\r) \phi_{Kk}(\r)
	& \approx \sum_{A=\{I,K\}}\sum_{\alpha\in A}
	C_{Ii, Kk}^{A\alpha} P_{A\alpha}(\r) \nonumber \\
  &= \suml_{\alpha\in I} C_{Ii, Kk}^{I\alpha} P_{I\alpha}(\r)
	+ \suml_{\alpha\in K} C_{Kk, Ii}^{K\alpha} P_{K\alpha}(\r)\,
	\label{eq:LRI}
\end{align}
where, similar to the AO case (eq~\eqref{eq:AO}),
\begin{equation}
   P_{A\alpha}(\r) \defeq P_{\aR{A}(\R_A)\alpha}(\r)=P_\alpha(\r-{\bm \tau}_{\aR{A}} -\R_A)
   \label{eq:ABF}
\end{equation}
and $C_{Ii, Kk}^{A\alpha}$ with $A=\{I,K\}$ are the two-center, three-index expansion coefficients.
The lower (Roman) and upper (Greek) indices of the $C$ coefficients denote the AOs and ABFs, respectively.
The essence of eq~\eqref{eq:LRI} is that the ABFs are restricted either on the atom $I$ or the atom $K$,
on which the two AOs are centering.
It should be noted that, the expansion coefficients $C$,
although carrying three orbital indices,
can be obtained in terms of two-center integrals \cite{Ihrig/etal:2015,Lin/Ren/He:2020},
which may be evaluated rather efficiently using the  algorithm developed by Talman\cite{Talman78-SBT,Talman84-4center,Talman03-MCI} for NAOs.

Using eq~\eqref{eq:LRI},
the computational expression of the HFX in eq~\eqref{eq:HFX_ERIs} naturally splits into four pieces,
\begin{align}
	\H_{Ii,Jj} \approx \sum_{K,L}\sum_{k \in K,l \in L}
	 & \left[ \sum_{\alpha \in I, \beta \in J} C_{Ii,Kk}^{I\alpha}(P_{I\alpha}|P_{J\beta})C_{Jj,Ll}^{J\beta} +
	     \sum_{\alpha \in I, \beta \in L}C_{Ii,Kk}^{I\alpha}(P_{I\alpha}|P_{L\beta})C_{Jj,Ll}^{L\beta} \right. \nonumber \\
	  & \left. \sum_{\alpha \in K, \beta \in J}C_{Ii,Kk}^{K\alpha}(P_{K\alpha}|P_{J\beta})C_{Jj,Ll}^{J\beta} +
       \sum_{\alpha \in K, \beta \in L}C_{Ii,Kk}^{K\alpha}(P_{K\alpha}|P_{L\beta})C_{Jj,Ll}^{L\beta}\right] D_{Kk,Ll}\,
\label{eq:HFX_LRI}
\end{align}
where $(P_{A\alpha}|P_{B\beta})$ with $A=\{I,K\}$ and $B=\{J,L\}$ are the (screened) Coulomb integrals between two ABFs,
defined by eq~\eqref{eq:coulomb_integrals}.
For notational convenience, below we also denote the (screened) Coulomb integrals as
$V_{A\alpha,B\beta} \defeq (P_{A\alpha}|P_{B\beta})$.

In order to present our algorithm in a clear way, we re-express eq~\eqref{eq:HFX_LRI} as
\begin{equation}
	\H_{Ii,Jj} \approx \sum_{K,L}
	\left[
		\Ht_{\Patom{I}i,K|\Patom{J}j,L} + \Ht_{\Patom{I}i,K|Jj,\Patom{L}}
		+ \Ht_{Ii,\Patom{K}|\Patom{J}j,L} + \Ht_{Ii,\Patom{K}|Jj,\Patom{L}}
	\right]
\label{eq:HFX_split}\end{equation}
where each of four terms in eq~\eqref{eq:HFX_split} stems from a corresponding one in eq~\eqref{eq:HFX_LRI}, e.g.,
\begin{align}
    \Ht_{\Patom{I}i,K|\Patom{J}j,L}&\defeq\sum_{k \in K,l \in L}
	  \sum_{\alpha \in I, \beta \in J} C_{Ii,Kk}^{I\alpha} V_{I\alpha,J\beta} C_{Jj,Ll}^{J\beta} D_{Kk,Ll} \nonumber \\
	  &= \suml_{k\in K, l\in L} [\phi_{\Patom{I}i} \phi_{Kk} | \phi_{\Patom{J}j} \phi_{Ll}] D_{Kk,Ll} \,
	  \label{eq:HFX_individual_term}
\end{align}
with the underlined atomic indices highlighting those two atoms on which the ABFs are located.
Furthermore, for the $\Ht$ objects,
the atoms $K$ and $L$ don't carry AO indices $k,l$ since they have been contracted out with the density matrix.
In eq~\eqref{eq:HFX_individual_term},
we have introduced partial two-electron Coulomb integrals denoted by square brackets,
\begin{equation}
	[\phi_{\Patom{A}a} \phi_{Ff} | \phi_{\Patom{B}b} \phi_{Gg}]
	\defeq \suml_{\alpha \in A, \beta \in B} C_{Aa,Ff}^{A\alpha} V_{A\alpha,B\beta} C_{Bb,Gg}^{B\beta} \,
	\label{eq:partial_integrals}
\end{equation}
where $A, B, F, G$ are atomic indices and $a,b,f,g$ are AOs centering on these atoms.
Again, the underscored indices for atoms $A$ and $B$ indicated these are the atoms where the ABFs are located.
Note that exchanging the positions of $A$ and $F$,
or $B$ and $G$ in the square bracket does not change the value of the integral, namely,
$[\phi_{\Patom{A}a} \phi_{Ff} | \phi_{\Patom{B}b} \phi_{Gg}]=[\phi_{Ff} \phi_{\Patom{A}a} | \phi_{\Patom{B}b} \phi_{Gg}]=
[\phi_{\Patom{A}a} \phi_{Ff}  | \phi_{Gg}\phi_{\Patom{B}b}]=[\phi_{Ff} \phi_{\Patom{A}a} | \phi_{Gg}\phi_{\Patom{B}b}]$.
In this work, the capital letters $I,J$ are reserved for the atoms to which the AO indices of the HFX matrix belong, 
$K,L$ for the atoms to the AO indices of the density matrix belong, whereas $A,B$ are referred to the atoms on which
the ABFs are sitting, and $F,G$ for the atoms without ABFs. According to the definition given by eq~\eqref{eq:partial_integrals},
one can easily see that the full ERIs are given by the sum of four partial ERIs,
\begin{equation}
   (\phi_{Ii} \phi_{Kk} | \phi_{Jj} \phi_{Ll})
   \approx [\phi_{\Patom{I}i} \phi_{Kk} | \phi_{\Patom{J}j} \phi_{Ll}]
   + [\phi_{\Patom{I}i} \phi_{Kk} | \phi_{Jj}\phi_{\Patom{L}l} ]
   + [\phi_{Ii} \phi_{\Patom{K}k} | \phi_{\Patom{J}j} \phi_{Ll}]
   + [\phi_{Ii} \phi_{\Patom{K}k} | \phi_{Jj}\phi_{\Patom{L}l} ] \, .
   \label{eq:ERI_split}
\end{equation}
However, as detailed below, in our practical implementations,
we never need to form the full ERIs according to eq~\eqref{eq:ERI_split}.
Rather, the $\Ht$ objects, given by the contraction of the partial ERIs with the density matrix,
are used to build the HFX matrix.
This choice is made for the sake for designing highly efficient parallel algorithms.

Equations \eqref{eq:HFX_split}-\eqref{eq:partial_integrals} serve as the mathematical foundation for one
to design efficient practical algorithms to evaluate the HFX matrix.
We remark that the error incurred by LRI, as given by eq~\eqref{eq:LRI},
can be made sufficiently small, provided that suitable auxiliary basis sets can be constructed.
The accuracy aspect of our HDF implementation based on
LRI has been benchmarked in a previous paper \cite{Lin/Ren/He:2020}.
In the present paper, we focus on the implementation details,
as well as the efficiency and the scaling behavior of the implementation.

\section{\label{sec:imple}Implementation Details}
As mentioned above,
the key underlying equations behind our implementation are eqs~\eqref{eq:HFX_split}-\eqref{eq:partial_integrals},
on which our algorithm for building the HFX matrix is based.
To achieve good efficiency and scalability, specific details need to be considered,
such as how to design the loop structure,
how to screen out the ERIs below a pre-given threshold,
and how to distribute the computational load and memory storage over the MPI (Message Passing Interface\cite{MPI}) tasks.
These details will be discussed in the present section.



\subsection{Loop structure for evaluating the HFX matrix \label{sec:Hexx}}

The loop structure for a straightforward evaluation of the HFX matrix is presented in Algorithm~\ref{lst:Hexx_4_IJ}.
In this algorithm, one first loops over the atomic indices of the HFX matrix,
i.e., the atom pair \pair{I}{J}.
Here the translational symmetry of periodic systems can be taken into account by restricting the atom $I$ within the
central unit cell (i.e., setting $\R_I=\0$),
while allowing the atom $J$ to be placed in the entire supercell.
For each atom pair \pair{I}{J},
one goes through all the atoms $K$ in the neighborhood of the atom $I$ (denoted as $K\in \Neighbour{I}$),
and  all the atoms $L$ in the neighborhood of $J$ (i.e., $L \in \Neighbour{J}$).
Within the loop over atoms $K$ and $L$, 
one can form all the partial ERIs belonging to these four atoms, namely,
$[\phi_{\Patom{I}i} \phi_{Kk} | \phi_{\Patom{J}j} \phi_{Ll}]$,
$[\phi_{\Patom{I}i} \phi_{Kk} | \phi_{Jj} \phi_{\Patom{L}l}]$,
$[\phi_{Ii} \phi_{\Patom{K}k} | \phi_{\Patom{J}j} \phi_{Ll}]$,
$[\phi_{Ii} \phi_{\Patom{K}k} | \phi_{Jj} \phi_{\Patom{L}l}]$.
However, one may choose not to form the full ERIs according to eq~\eqref{eq:ERI_split} here.
This is because, in parallel computing,
these partial ERIs are evaluated and stored on different processes, 
and forming the full ERIs requires a significant amount of communication.
Instead, they are contracted locally with the density matrix,
ending up with the much smaller $\Ht$ objects, as illustrated in Algorithm~\ref{lst:Hexx_4_IJ}.
Adding up these $\Ht$ objects for all \pair{K}{L} atom pairs, one obtains the full desired HFX matrix.

Such a straightforward calculation flow works, but is not an efficient one.
In fact, in Algorithm~\ref{lst:Hexx_4_IJ},
each independent partial ERI has been computed four times, which is a big waste.
For example, as noted above,
the partial ERI $[\phi_{\Patom{I}i} \phi_{Kk} | \phi_{\Patom{J}j} \phi_{Ll}]$
is equal to $[\phi_{Kk}\phi_{\Patom{I}i} | \phi_{\Patom{J}j} \phi_{Ll}]$ due to the presence of permutation symmetry,
but both need to be computed in Algorithm~\ref{lst:Hexx_4_IJ}.
The same happens when one swaps the $\Patom{J}$ and $L$ indices.
This is because these partial ERIs cannot be stored in memory at the same time,
and hence they are computed on the fly within the \pair{I}{J} loop.
As a consequence, only partial ERIs with a fixed \pair{I}{J} pair are available at a time.
Thus for $I\ne K$, $[\phi_{Kk} \phi_{\Patom{I}i} | \phi_{\Patom{J}j} \phi_{Ll}]$ always needs to computed,
since its equivalent $[\phi_{\Patom{I}i} \phi_{Kk} | \phi_{\Patom{J}j} \phi_{Ll}]$ is not available any,
even if it has been computed before.

To overcome this redundancy issue,
we switch to a revised loop structure as presented in Algorithm~\ref{lst:Hexx_4_AB}.
A key feature of Algorithm~\ref{lst:Hexx_4_AB} is that each
independent partial ERI is computed only once.
To this end, instead of first looping over the indices of the HFX matrix,
i.e., the \pair{I}{J} pair, we loop over the \pair{A}{B} pair, on which the ABFs are located.
(Again, the translation symmetry can be accounted by restricting $A$ in the central unit cell,
while allowing $B$ to be located in the entire supercell.)
Within the  \pair{A}{B} loop, one computes and stores the partial ERIs
$[\phi_{\Patom{A}a} \phi_{Ff} | \phi_{\Patom{B}b} \phi_{Gg}]$ for all $F \in \Neighbour{A}$ and $G \in \Neighbour{B}$.
To maximize the usage of the partial ERIs that are present at a time,
we contract the fourth-rank tensor $[\phi_{\Patom{A}a} \phi_{Ff} | \phi_{\Patom{B}b} \phi_{Gg}]$
for a given set of atoms $A,B,F,G$ with the density matrix  $D_{Ff,Gg}$, $D_{Ff,Bb}$, $D_{Aa,Gg}$, and $D_{Aa,Bb}$,
ending up with contributions to different sectors of the HFX matrix,
as illustrated in Algorithm~\ref{lst:Hexx_4_AB}.
In this algorithm, the permutation symmetry of the ERIs is naturally incorporated,
and each independent partial ERI is computed only once.
The trick here is that, with the new loop structure,
we can contract the partial ERIs with
different subblocks of the density matrix originating from different pairs of atoms.
When the loop over \pair{A}{B} is complete, the full HFX matrix is obtained.
This redesign of the loop structure speeds up the calculations by roughly a factor of 4.

\begin{algorithm}
	\caption{Principal loop structure based on \pair{I}{J} atomic pairs }
	\label{lst:Hexx_4_IJ}
	\begin{algorithmic}[1]
		\ForAll{$\pair{I}{J}$}
			\ForAll{$K \in \Neighbour{I}$, $L \in \Neighbour{J}$}
				\State Calculate $[\phi_{\Patom{I}i} \phi_{Kk} | \phi_{\Patom{J}j} \phi_{Ll}]$		
				\State Calculate $[\phi_{\Patom{I}i} \phi_{Kk} | \phi_{Jj} \phi_{\Patom{L}l}]$		
				\State Calculate $[\phi_{Ii} \phi_{\Patom{K}k} | \phi_{\Patom{J}j} \phi_{Ll}]$		
				\State Calculate $[\phi_{Ii} \phi_{\Patom{K}k} | \phi_{Jj} \phi_{\Patom{L}l}]$		
				\State $\Ht_{\Patom{I}i, K|\Patom{J}j, L} = \suml_{k\in K, l\in L} [\phi_{\Patom{I}i} \phi_{Kk} | \phi_{\Patom{J}j} \phi_{Ll}] * D_{Kk,Ll}$
				\State $\Ht_{\Patom{I}i, K|Jj, \Patom{L}} = \suml_{k\in K, l\in L} [\phi_{\Patom{I}i} \phi_{Kk} | \phi_{Jj} \phi_{\Patom{L}l}] * D_{Kk,Ll}$
				\State $\Ht_{Ii, \Patom{K}|\Patom{J}j, L} = \suml_{k\in K, l\in L} [\phi_{Ii} \phi_{\Patom{K}k} | \phi_{\Patom{J}j} \phi_{Ll}] * D_{Kk,Ll}$
				\State $\Ht_{Ii, \Patom{K}|Jj, \Patom{L}} = \suml_{k\in K, l\in L} [\phi_{Ii} \phi_{\Patom{K}k} | \phi_{Jj} \phi_{\Patom{L}l}] * D_{Kk,Ll}$
				\State $\H_{Ii,Jj} += \Ht_{\Patom{I}i, K|\Patom{J}j, L}$		
				\State $\H_{Ii,Jj} += \Ht_{\Patom{I}i, K|Jj, \Patom{L}}$		
				\State $\H_{Ii,Jj} += \Ht_{Ii, \Patom{K}|\Patom{J}j, L}$		
				\State $\H_{Ii,Jj} += \Ht_{Ii, \Patom{K}|Jj, \Patom{L}}$		
			\EndFor
		\EndFor
	\end{algorithmic}
\end{algorithm}

\begin{algorithm}
	\caption{Principal loop structure based on \pair{A}{B} atomic pairs}
	\label{lst:Hexx_4_AB}
	\begin{algorithmic}[1]
		\ForAll{$\pair{A}{B}$}
			\ForAll{$F \in \Neighbour{A}$, $G \in \Neighbour{B}$}
				\State Calculate $[\phi_{\Patom{A}a} \phi_{Ff} | \phi_{\Patom{B}b} \phi_{Gg}]$
				\State $\Ht_{\Patom{A}a, F  | \Patom{B}b, G } = \suml_{f\in F, g\in G} [\phi_{\Patom{A}a} \phi_{Ff} | \phi_{\Patom{B}b} \phi_{Gg}] * D_{Ff,Gg}$
				\State $\Ht_{\Patom{A}a, F  | \Patom{B} , Gg} = \suml_{f\in F, b\in B} [\phi_{\Patom{A}a} \phi_{Ff} | \phi_{\Patom{B}b} \phi_{Gg}] * D_{Ff,Bb}$
				\State $\Ht_{\Patom{A} , Ff | \Patom{B}b, G } = \suml_{a\in A, g\in G} [\phi_{\Patom{A}a} \phi_{Ff} | \phi_{\Patom{B}b} \phi_{Gg}] * D_{Aa,Gg}$
				\State $\Ht_{\Patom{A} , Ff | \Patom{B} , Gg} = \suml_{a\in A, b\in B} [\phi_{\Patom{A}a} \phi_{Ff} | \phi_{\Patom{B}b} \phi_{Gg}] * D_{Aa,Bb}$
				\State $\H_{Aa, Bb} += \Ht_{\Patom{A}a, F  | \Patom{B}b, G }$ 	
				\State $\H_{Aa, Gg} += \Ht_{\Patom{A}a, F  | \Patom{B} , Gg}$ 	
				\State $\H_{Ff, Bb} += \Ht_{\Patom{A} , Ff | \Patom{B}b, G }$ 	
				\State $\H_{Ff, Gg} += \Ht_{\Patom{A} , Ff | \Patom{B} , Gg}$ 	
			\EndFor
		\EndFor
	\end{algorithmic}
\end{algorithm}



In addition to improved efficiency,
Algorithm~\ref{lst:Hexx_4_AB} is also advantageous for memory usage in case of parallel computing.
To understand this point, we note that the atom pairs at the outermost loop are distributed over the MPI processes.
Specifically, in Algorithm~\ref{lst:Hexx_4_IJ}, the \pair{I}{J} pairs are distributed,
and for a given \pair{I}{J} pair, the information for $\forall K \in \Neighbour{I}$ and $\forall L \in \Neighbour{J}$ is stored locally.
These include subblocks of both the Coulomb matrix $V_{K\mu,L\nu}$
and the density matrix $D_{Kk,Ll}$ originating from $K$ and $L$ atoms.
Since the same atom pairs \pair{K}{L} can be neighbors of different \pair{I}{J} pairs,
there will be a duplication of \pair{K}{L} atom pairs,
and hence a duplication of $V_{K\alpha,L\beta}$ and $D_{Kk,Ll}$ matrices over MPI processes,
but there is no duplication of the $\H_{Ii,Jj}$ matrix.
Following a similar line of reasoning, in Algorithm~\ref{lst:Hexx_4_AB},
$D_{Ff,Gg}$ and $\H_{Ff,Gg}$ matrices are duplicated, but there is no duplication of $V_{A\alpha,B\beta}$ matrix.
Since the size of ABFs is several times larger than that of the AOs,
the size of the $V$ is much larger than that of the $\H$ matrix.
Thus Algorithm~\ref{lst:Hexx_4_AB} consumes less memory than
Algorithm~\ref{lst:Hexx_4_IJ} and this is the second advantage of Algorithm~\ref{lst:Hexx_4_AB}.

\begin{figure}[!htbp]
	\centering
	\includegraphics[width=0.5\textwidth]{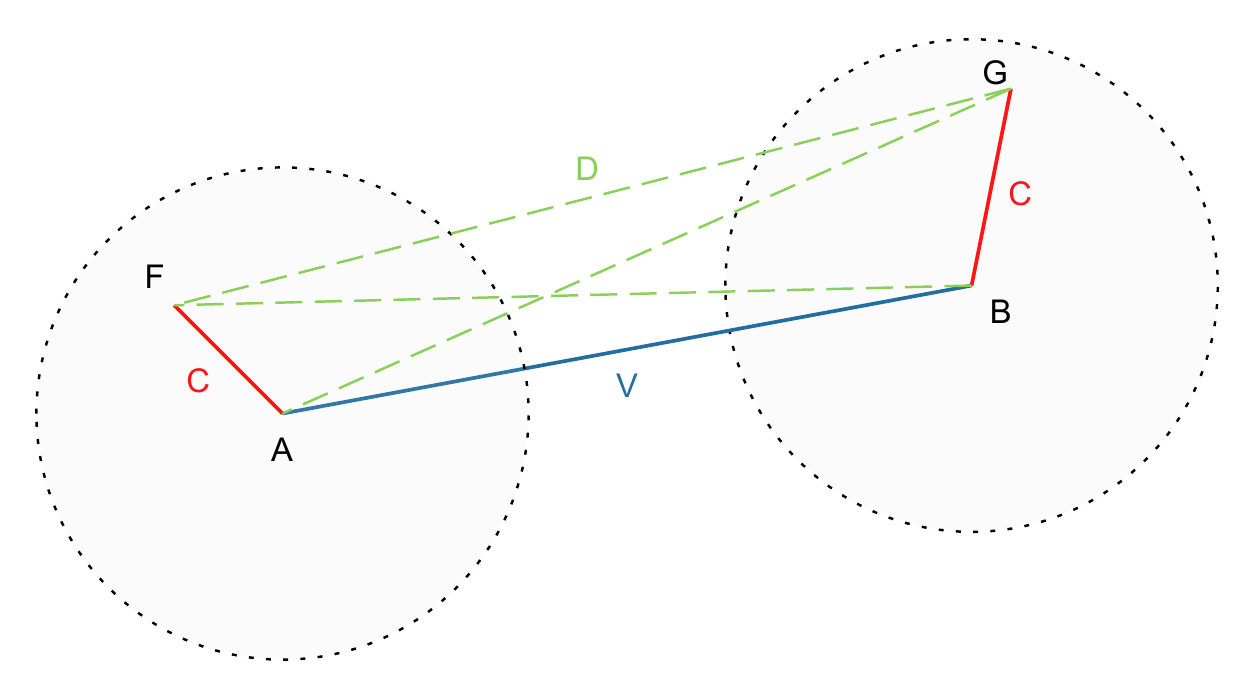}
	\caption{
		Pictorial illustration of Algorithm~\ref{lst:Hexx_4_AB}.
		The four dots denote four atoms $A,B,F$ and $G$,
		where $A$ and $B$ are the atoms where the ABFs are located,
		and hence they are connected by a $V$ (interaction) line.
		The atoms $F$ and $G$ sit in
		the neighborhood of $A$ and $B$, respectively,
		represented by two circles centered at $A$ and $B$.
		The atoms $A$ and $F$,
		as well as $B$ and $G$ are connected by the $C$ (expansion coefficient) lines.
		The green dash lines denote the density matrix between two atoms.
		For periodic systems the atom $A$ is restricted to the central unit cell, whereas the
		atom $B$ is located in the entire supercell. 
	}
	\label{fig:CVCD}
\end{figure}

A pictorial illustration of Algorithm~\ref{lst:Hexx_4_AB} is presented in Fig.~\ref{fig:CVCD}, where
the atoms are denoted by dots and the expansion coefficients $C$, the Coulomb matrix $V$, and the density matrix $D$
are represented by lines. Finally we remark that the flowcharts presented in Algorithm \ref{lst:Hexx_4_IJ} and 
\ref{lst:Hexx_4_AB} are mainly used to elucidate our choice of the major loop structure of our implementation. 
In practical implementation, it is in fact not necessary (and too expensive) to explicitly form the partial ERIs
$[\phi_{\Patom{A}a} \phi_{Ff} | \phi_{\Patom{B}b} \phi_{Gg}]$. As shown in Appendix~\ref{sec:matrix_multiply}, 
the most efficient way to proceed is to first contract the Coulomb matrix $V$ with the expansion coefficients $C$,
and then multiply the resultant quantity with the density matrix $D$. At a final step, the HFX matrix by
multiply $C$ with the product of $(VC)$ and $D$. Namely, $\H \sim C((VC)D)$. An illustration of this refined procedure adopted in
our practical implementation is given in Algorithm~\ref{Algo3_AB}.

\begin{algorithm}
	\caption{Refined loop structure adopted in practical implementation. Here $Y=V*C$, and $T=Y*D$ are temporary
	rank-3 tensors}
	\label{Algo3_AB}
	\begin{algorithmic}[1]
		\ForAll{$\pair{A}{B}$}
			\ForAll{$F \in \Neighbour{A}$, $G \in \Neighbour{B}$}
				\State $Y_{Bb,Gg}^{A\alpha} = \suml_{\beta\in B} V_{A\alpha,B\beta} * C_{Bb,Gg}^{B\beta}$
				\State $T_{Ff,Bb}^{A\alpha} = \suml_{g\in G} Y_{Bb,Gg}^{A\alpha} * D_{Ff,Gg}$
				\State $T_{Ff,Gg}^{A\alpha} = \suml_{b\in B} Y_{Bb,Gg}^{A\alpha} * D_{Ff,Bb}$
				\State $T_{Aa,Bb}^{A\alpha} = \suml_{g\in G} Y_{Bb,Gg}^{A\alpha} * D_{Aa,Gg}$
				\State $T_{Aa,Gg}^{A\alpha} = \suml_{b\in B} Y_{Bb,Gg}^{A\alpha} * D_{Aa,Bb}$
				\State $\Ht_{\Patom{A}a,F |\Patom{B}b,G } = \suml_{f\in F} \suml_{\alpha\in A} C_{Aa,Ff}^{A\alpha} * T_{Ff,Bb}^{A\alpha}$
				\State $\Ht_{\Patom{A}a,F |\Patom{B}, Gg} = \suml_{f\in F} \suml_{\alpha\in A} C_{Aa,Ff}^{A\alpha} * T_{Ff,Gg}^{A\alpha}$
				\State $\Ht_{\Patom{A}, Ff|\Patom{B}b,G } = \suml_{a\in A} \suml_{\alpha\in A} C_{Aa,Ff}^{A\alpha} * T_{Aa,Bb}^{A\alpha}$
				\State $\Ht_{\Patom{A}, Ff|\Patom{B}, Gh} = \suml_{a\in A} \suml_{\alpha\in A} C_{Aa,Ff}^{A\alpha} * T_{Aa,Gg}^{A\alpha}$
				\State $\H_{Aa,Bb} += \Ht_{\Patom{A}a,F |\Patom{B}b,G }$
				\State $\H_{Aa,Gg} += \Ht_{\Patom{A}a,F |\Patom{B} ,Gg}$
				\State $\H_{Ff,Bb} += \Ht_{\Patom{A} ,Ff|\Patom{B}b,G }$
				\State $\H_{Ff,Gg} += \Ht_{\Patom{A} ,Ff|\Patom{B} ,Gg}$
			\EndFor
		\EndFor
	\end{algorithmic}
\end{algorithm}



\subsection{Sparsity of matrices and pre-screening \label{sec:sparsity}}

As discussed above, the most time-consuming step in self-consistent HDF calculations is to
construct the HFX matrix.
The building blocks of the HFX matrix is the $\Ht$ objects,
introduced in eqs~\eqref{eq:HFX_split} and \eqref{eq:HFX_individual_term},
which are given schematically by $\Ht=CVCD$, where $C$, $V$, and $D$ denote, respectively,
the expansion coefficient matrix (eq~\eqref{eq:LRI}),
the (screened) Coulomb matrix, and the density matrix.
A straightforward evaluation of all $\Ht$ objects scales as $N_{\rm at}^4$,
where $N_{\rm at}$ is the number of atoms in the system.
However, due to the locality of the NAOs and, in case of insulators, the rapidly decaying behavior of the density matrix
in real space,
only the exchange interactions within a limited spatial range are needed.
In fact, a large portion of elements of the matrix $\Ht$,
as well as those of the $C$, $V$ and $D$ matrices are extremely small or even strictly zero.
Eliminating these small matrix elements has little effect on the obtained results,
but can save a lot of computation time and memory.
As will be shown below, by exploiting the sparsity of these matrices,
one can achieve linear scaling of the computational cost for evaluating the $\H$ matrices,
and hence a linear-scaling build of the HFX matrix.

\subsubsection{The matrix of the expansion coefficients $C$ \label{sec:sparsity_C}}

Let's first look at the $C$ matrix.
Because the NAOs have a finite cut-off radius,
their overlap and consequently the expansion coefficients $C_{Aa,Ff}^{A\alpha}$
become strictly zero if the distance between two atoms $A$ and $F$
is larger than the sum of the cut-off radii of AO basis functions $a$ and $f$, as illustrated in Fig.~\ref{fig:CVCD}.
When discussing Algorithm~\ref{lst:Hexx_4_IJ} and \ref{lst:Hexx_4_AB}
we already introduced the notation of neighboring atoms.
Here we shall define that the neighboring atoms $F$ of an atom $A$
are those for which the expansion coefficients $C_{Aa,Ff}^{A\alpha}$ are non-zero, i.e.,
$\Neighbour{A} \defeq \{ F | C_{Aa,Ff}^{A\alpha} \neq 0, \forall a\in A, f\in F, \alpha \in A \}$.
The matrix elements of $C$ originating from atom pairs that are not ``neighbors" are strictly zero and
are excluded from the outset of the calculations. In addition, there are  $C_{Aa,Ff}^{A\alpha}$ elements which are sufficiently small to be negligible, 
without affecting the obtained results. As detailed below, these can be filtered out to further save computation time and memory.

\subsubsection{The (screened) Coulomb matrix $V$}
\label{sec:coulomb_matrix_screening}
The matrix elements of the (screened) Coulomb interaction between the ABFs are calculated as,
\begin{equation}\begin{array}{rcl}
	V_{A\alpha,B\beta}
	&=& (P_{A\alpha} | P_{B\beta})			\\
	&=& \braket{P_{A\alpha} | Q_{B\beta}}
\end{array}\end{equation}
where $Q_{B\beta}(\r) = \int v(\r-\r') P_{B\beta}(\r') d\r'$, and $\braket{f|g}$ denotes the overlap integral
between two functions $f(\r)$ and $g(\r)$.


For the bare Coulomb potential $v(\r,\r') = 1/|\r-\r'|$,
$Q_\beta(\r)$ is has a rather long range and the sparsity of the $V$ matrix is low.
In fact, in this case certain elements in
the Coulomb matrix become diverging at the $\Gamma$ point when Fourier transformed to $\k$ space.
Well-established procedures exist to deal with this so-called $\Gamma$-point singularity
\cite{Gygi/Baldereschi:1986,Spencer/Alavi:2008,Levchenko/etal:2015}
and we don't discuss this issue in the present paper. For screened HDFs such as HSE, 
one employs the short-ranged (screened) Coulomb interaction
$v(\r,\r') = \text{erfc}(\mu|\r-\r|)/|\r-\r|$, and hence $Q_\beta(\r)$ is short ranged.
In this case, one can introduce a finite cut-off radius $R_Q$ for $Q_\beta(\r)$,
beyond which the elements of the (screened) Coulomb matrix become sufficiently small and can be 
neglected.
Of course the sparsity of the obtained $V$ matrix depends on the screening parameter $\mu$.


\subsubsection{The density matrix $D$ \label{sec:sparsity_D}}

The density matrix $D_{Kk,Ll}$ of an insulator has an exponential decay behavior
as a function of the distance between the two atoms
\cite{Kohn:1959,Cloizeaux/etal:1964,he2001exponential}.
Therefore, if two atoms are far apart,
the value of $D_{Kk,Ll}$ may be small enough to be negligible.
However, for metallic systems,
the density matrix only has a polynomial delay, and its sparsity is much reduced.
Similar to the case of $C$, the insignificant elements of $D$ are also filtered out, as detailed below.

\subsubsection{$\Ht$ matrices}

The $\Ht$ matrices are given by the product of $C$, $V$, and $D$, and hence their sparsity originates from these individual matrices.
To exploit the sparsity of these matrices and to make the best use of BLAS\cite{lawson1977basic,dongarra1988extended,dongarra1990algorithm}
matrix multiplications which are most suitable for dense matrices,
an optimal strategy is to organize and store the elements of the these matrices as sparsely distributed dense subblocks.
With this in mind, in our implementation the data structure of these four tensors
are all stored as aggregates of small block tensors.
For example, the global matrix $C_{Aa,Ff}^{A\alpha}$ can be represented as a supermatrix $C_{A,F}$, which itself
is a sparse matrix, while its non-zero elements are dense rank-3 tensors $C_{a,f}^{\alpha}$.
Similarly, 
the global $V$ and $D$ matrices can be seen as atom-pair based supermatrices $V_{A,B}$ and $D_{K,L}$ ($K=\{A,F\}$, $L=\{B,G\}$, 
see Algorithm~\ref{lst:Hexx_4_AB}), 
with each element of them being a rank-2 tensor $V_{\alpha,\beta}$ and $D_{k,l}$, respectively.
To exploit the sparsity of $C$ mentioned in \ref{sec:sparsity_C}, an upper limit
\begin{equation}
\overline{C_{A,F}} \defeq \max_{a\in A,f\in F,\alpha\in A} |C_{Aa,Ff}^{A\alpha}|
\end{equation}
for each block $C_{A,F}$ is introduced.
In practical implementation, one can introduce a finite threshold parameter $\varepsilon_C$.
If $\overline{C_{A,F}} \leq \varepsilon_C$, the entire block $C_{A,F}$ (i.e., the contribution from the \pair{A}{F} atom pair) is disregarded.
Similarly, an upper bound $\overline{D_{K,L}}$  can be defined for each block $D_{K,L}$, and the density matrix blocks with
$\overline{D_{K,L}}\leq\varepsilon_D$ are discarded. The filtering of negligible blocks of the $V$ matrix is controlled by
the cut-off radius $R_Q$. The influence of $\varepsilon_C$, $\varepsilon_D$, and $R_Q$
on the accuracy of the obtained results and the computational efficiency will be discussed in
Sec.~\ref{sec:result_sparsity}.

Due to the strict locality of the supermatrix $C_{A,F}$ ($C_{B,G}$),
the atom $F$ ($G$) is constrained to the neighbouring atoms of $A$ ($B$).
Furthermore, as long as either the Coulomb matrix (in case of screened HDFs ) 
or density matrix (in case of insulators) is short-ranged,
only atom pairs \pair{A}{B} within a certain range contribute to the $\Ht$ matrices.
In other words, for a given atom $A$,
the numbers of neighbouring atoms $B$, $F$ and $G$ contributing to the final HFX matrix are independent of system size,
which warrants a linear-scaling computational cost for evaluating the HFX matrix, at least for screened HDFs or
for insulating systems.

According to the formal relationship $\Ht \sim CVCD$, it is obvious that the sparsity of $\Ht$ comes from two aspects.
For one thing, the zeros of individual blocks $C_{A,F}$, $V_{A,B}$, and $D_{K,L}$ directly lead to the sparsity of 
the $\Ht$ matrices. For example, for a given set of atoms $A,B,F,G$, the matrix block $\Ht_{\Patom{A}a, F  | \Patom{B}b, G }$
is non-zero only if the four multiplier matrix blocks are all non-zero.
Another possibility is that, even if the elements of the component matrices exceed their own screening thresholds,
certain elements of the resultant product matrix $\Ht$ may still be negligibly small.
These small matrix elements can be efficiently screened out by making use of the Cauchy-Schwarz inequalities.
The detailed screening procedures will be discussed in Sec.~\ref{sec:cauchy} and Sec.~\ref{sec:schwarz}.


\subsection{The screening algorithm}

As discussed above, even after pre-screening the individual $C$, $V$, and $D$ matrices, there is still a portion of the 
$\Ht$ matrix elements being extremely small,
and can be safely neglected without affecting the obtained results. Thus, it would be highly desirable 
if the insignificant elements of the $\Ht$ matrices can be efficiently filtered out,
without actually calculating them. This is achieved by estimating the upper bounds of these matrix elements based on
the Cauchy-Schwarz inequality, as detailed below, and neglecting the elements below a pre-chosen threshold at
the early stage of the calculations. 

\subsubsection{Cauchy-Schwarz inequality matrix screening \label{sec:cauchy}}

The upper bounds of the matrix elements of $\Ht$ are estimated in terms of Cauchy-Schwarz inequality, according to which
the product of two matrices ${\cal A}$ and ${\cal B}$ satisfies
\begin{equation}
	|\tr[{\cal A} {\cal B}]| \leq \sqrt{\tr[{\cal A}^+{\cal A}]} \sqrt{\tr[{\cal B}^+{\cal B}]} = \|{\cal A}\| \|{\cal B}\| \,
\end{equation}
where $\|{\cal A}\| = \sqrt{ \suml_{ij} |a_{ij}|^2 }$ is the L2-norm of the matrix ${\cal A}$.
The Cauchy-Schwarz inequality can be extended straightforwardly to the multiplication of three and more matrices, e.g.,
\begin{equation}\begin{array}{cl}
	 & |\tr[{\cal A}{\cal B}{\cal C}]|		\\
	\leq& \sqrt{\tr[{\cal A}^+{\cal A}]} \sqrt{\tr[({\cal B}{\cal C})^+({\cal B}{\cal C})]}	\\
	=& \|{\cal A}\| \sqrt{\tr[({\cal B}^+{\cal B})({\cal C}{\cal C}^+)]}			\\
	\leq& \|{\cal A}\| \sqrt{\|{\cal B}^+{\cal B}\|} \sqrt{\|{\cal C}{\cal C}^+\|}
\end{array}\end{equation}
and
\begin{equation}\begin{array}{cl}
	 & |\tr[{\cal A}{\cal B}{\cal C}{\cal D}]|		\\
	\leq& \sqrt{\tr[({\cal A}{\cal B})^+({\cal A}{\cal B})]} \sqrt{\tr[({\cal C}{\cal D})^+({\cal C}{\cal D})]}		\\
	=& \sqrt{\tr[({\cal A}^+{\cal A})({\cal B}{\cal B}^+)]} \sqrt{\tr[({\cal C}^+{\cal C})({\cal D}{\cal D}^+)]}			\\
	\leq& \sqrt{\|{\cal A}^+{\cal A}\|} \sqrt{\|{\cal B}{\cal B}^+\|} \sqrt{\|{\cal C}^+{\cal C}\|} \sqrt{\|{\cal D}{\cal D}^+\|}	\\
\end{array}\end{equation}

Now we can apply the Cauchy-Schwarz inequality to eq~\eqref{eq:HFX_individual_term} for a fixed set of atoms $I,J,K,L$, and obtain 
(symbolically)
\begin{equation}\begin{array}{rcll}
   \Ht_{\Patom{I}i,K|\Patom{J}j,L}&= &\sum_{k \in K,l \in L}
	  \sum_{\alpha \in I, \beta \in J} & C_{Ii,Kk}^{I\alpha} V_{I\alpha,J\beta} C_{Jj,Ll}^{J\beta} D_{Kk,Ll} \\
	&=& \tr[ C_i V C_j D ]
	&\left( \leq \sqrt{\|C_iC_i^+\|} \sqrt{\|VV^+\|} \sqrt{\|C_jC_j^+\|} \sqrt{\|DD^+\|} \right)		\\
	&=& \tr[ C_i (V C_j) D ]
	&\left( \leq \sqrt{\|C_iC_i^+\|} \|V C_j\| \sqrt{\|DD^+\|} \right)					\\
	&=& \tr[ C_i ((V C_j) D) ]
	&\left( \leq \|C_i\| \|(V C_j) D\| \right)									
\end{array}
\label{eq:CS_upper_bounds}
\end{equation}
where $V=V_{I,J}$ and $D=D_{K,L}$ here are blocks of the Coulomb matrix and density matrix originating from the atomic
pair \pair{I}{J} and \pair{K}{L}, respectively.
$C_i = C_{Ii,K}^{I}$
is also a rank-2 tensor representing a sector of the triple expansion coefficients on
atom pair \pair{I}{K} (i.e., $C_{I,K}$ introduced above) with fixed AO basis function $i$ 
(i.e., the rank-3 block tensor $C_{I,K}$ with fixed $i$).  
The crucial part in eq~\eqref{eq:CS_upper_bounds} is the different estimated upper bounds given in parenthesis for the same quantity. 
These different estimations can be utilized to filter out the insignificant elements at different stages of the calculation,
all together leading to a highly efficient screening algorithm.

As mentioned above and explained in Appendix~\ref{sec:matrix_multiply}, the best order of matrix multiplication 
to obtain $\Ht$ is given by the last line of eq~\eqref{eq:CS_upper_bounds}. The pseudocode of
the accordingly designed screening algorithm for evaluating $\Ht$ is illustrated in Algorithm~\ref{lst:cauchy}. 
In this algorithm, the actual working procedure goes as follows. We first calculate and store all needed quantities --
$\sqrt{\|C_iC_i^+\|}$, $\|C_i\|$, $\sqrt{\|VV^+\|}$ and $\sqrt{\|DD^+\|}$ in advance. 
During the actual calculation process of $\Ht$ matrices, one further evaluate quantities $\|V C_j\|$ and
$\|(V C_j) D\|$. The three upper bounds listed in eq~\eqref{eq:CS_upper_bounds} are calculated at appropriate
locations within the calculation loops, and compared to a pre-chosen threshold $\varepsilon_{\text{CS-matrix}}$.
If any of the three upper bounds is below the threshold, the corresponding elements of $\Ht$ matrices are set to zero, without
actually computing its value. In Sec.~\ref{sec:result_cs}, we will present benchmark results regarding the influence of the threshold
$\varepsilon_{\text{CS-matrix}}$ on accuracy and computation cost.

\begin{algorithm}
	\caption{Cauchy-Schwarz (CS) inequality matrix screening with a pre-chosen threshold $\varepsilon_{\text{CS-matrix}}$}
	\label{lst:cauchy}
	\begin{algorithmic}[1]
		\If{$\sqrt{\|C_iC_i^+\|} \sqrt{\|VV^+\|} \sqrt{\|C_jC_j^+\|} \sqrt{\|DD^+\|} > \varepsilon_{\text{CS-matrix}}$}
			\State Calculate $V C_j$ and $\|V C_j\|$
			\If{$\sqrt{\|C_iC_i^+\|} \|V C_j\| \sqrt{\|DD^+\|} > \varepsilon_{\text{CS-matrix}}$}
				\State Calculate $(V C_j) D$ and $\|(V C_j) D\|$
				\If{$\|C_i\| \|((V C_j) D)\| > \varepsilon_{\text{CS-matrix}}$}
					\State Calculate $\tr[C_i ((V C_j) D)]$
				\EndIf
			\EndIf
		\EndIf
	\end{algorithmic}
\end{algorithm}

\subsubsection{Cauchy-Schwarz inequality ERI screening \label{sec:schwarz}}

In addition to the screening criteria applied in Algorithm~\ref{lst:cauchy}, one can apply one more
criterion directly based on the full ERIs. The Cauchy-Schwarz inequality can again
be applied to estimate the upper bound of each ERI quickly\cite{haser1989improvements},
\begin{equation}
	(\phi_{Aa}\phi_{Ff}|\phi_{Bb}\phi_{Gg}) \leq
	\sqrt{(\phi_{Aa}\phi_{Ff}|\phi_{Aa}\phi_{Ff})}
	\sqrt{(\phi_{Bb}\phi_{Gg}|\phi_{Bb}\phi_{Gg})} \, .
\label{eq:CS-ERI}\end{equation}
In our implementation, we compute all the ``diagonal" ERIs $(\phi_{Aa}\phi_{Fg}|\phi_{Aa}\phi_{Fg})$
and determine 
\begin{equation}
\overline{(\phi_{A}\phi_{F}|\phi_{A}\phi_{F})} \defeq \max_{a\in A, f\in F} (\phi_{Aa}\phi_{Ff}|\phi_{Aa}\phi_{Ff})
\end{equation}
beforehand, which can be done with relative ease since the number of these scales linearly with the system size.
Now, the upper bound of each ERI can be found easily, and those below a given threshold $\varepsilon_{\text{CS-ERI}}$ can be
disregarded without actually calculating them.

In \textsc{ABACUS}, we actually only calculate (implicitly) the partial ERIs $[\phi_{\Patom{A}a}\phi_{Ff}|\phi_{\Patom{B}b}\phi_{Gg}]$
and its three variant form.
However, this does not affect the use of Cauchy-Schwarz inequality screening.
Before calculating $[\phi_{\Patom{A}a}\phi_{Ff}|\phi_{\Patom{B}b}\phi_{Gg}]$
we first use \eqref{eq:CS-ERI} to check
whether $(\phi_{Aa}\phi_{Fa}|\phi_{Bb}\phi_{Gg})$ is needed or not.
If not, then it's unnecessary to calculate $[\phi_{\Patom{A}a}\phi_{Ff}|\phi_{\Patom{B}b}\phi_{Gg}]$ and other
three partial ERIs which form the full ERI (eq.~\eqref{eq:ERI_split}). 



Combining the ``matrix product screening" as discussed in Sec.~\ref{sec:cauchy} and the ERI-based screening discussed above,
the entire flowchart of for evaluating the HFX matrix is presented in Algorithm~\ref{lst:exx_program}. At a given atomic
structure, the maximal ``diagonal"
ERI $(\phi_{A}\phi_{F}|\phi_{A}\phi_{F})$ is first determined for each atomic pair. They are then used to filter out
the atomic quartet whose ERIs are below a threshold $\varepsilon_{\text{CS-ERI}}$, at the beginning of
the HFX calculation. The ``matrix product screening" is only
applied for those atomic sets $\{A,B,F,G\}$ which passed the ERI-based screening.

\begin{algorithm*}
	\caption{Flowchart of the HFX Evaluation Program}
	\label{lst:exx_program}
	\begin{algorithmic}[1]

		\Function{Initialize}{}
		\Comment{Perform once for all}
			\State Construct ABFs
			\State Parallel task distribution
		\EndFunction

		\Function{Calculate $C$ and $V$}{}
		\Comment{Perform at each step of ionic motions}
			\State Calculate $V$ for atomic pairs \pair{A}{B} with distance $< R_Q$
			\State Calculate $C$ for blocks with $\overline{C_{AF}} > \varepsilon_C$
			\State Calculate $(\phi_{Aa}\phi_{Ff}|\phi_{Aa}\phi_{Ff})$ and determine $\overline{(\phi_{A}\phi_{F}|\phi_{A}\phi_{F})}$  for CS-ERI
			\State Calculate $\|C\|$, $\sqrt{\|C C^+\|}$ and $\sqrt{\|V V^+\|}$ for CS-matrix
		\EndFunction

		\Function{Calculate exact-exchange}{}
		\Comment{Perform at each electronic step}
			\State Transmit $D$ for blocks with $\overline{D_{KL}} > \varepsilon_D$
			\State Calculate $\sqrt{\|D D^+\|}$ for CS-matrix
			\State Calculate $\H$
			\State Calculate $E^{\textnormal{X}}$
			\State Transmit $\H$
		\EndFunction

		\Function{Calculate $\H$}{}
			\ForAll{\pair{A}{B}}
				\ForAll{$F \in \Neighbour{A}$, $G \in \Neighbour{B}$}
					\If{$ \overline{(\phi_{A}\phi_{F}|\phi_{A}\phi_{F})} \overline{(\phi_{B}\phi_{G}|\phi_{B}\phi_{G})} > \varepsilon_{\text{CS-ERI}} $}
						\If{$\sqrt{\|C C^+\|} \sqrt{\|V V^+\|} \sqrt{\|C C^+\|} \sqrt{\|D D^+\|} > \varepsilon_{\textnormal{CS-matrix}}$}
							\State Calculate $VC$ and $\|VC\|$
							\If{$\sqrt{\|C C^+\|} \|VC\| \sqrt{\|D D^+\|} > \varepsilon_{\text{CS-matrix}}$}
								\State Calculate $(VC)D$ and $\|(VC)D\|$
								\If{$\|C\| \|(VC)D\| > \varepsilon_{\text{CS-matrix}}$}
									\State Calculate $C((VC)D)$
								\EndIf
							\EndIf
						\EndIf
					\EndIf
				\EndFor
			\EndFor
		\EndFunction

	\end{algorithmic}
\end{algorithm*}

\subsection{Parallelization of the algorithm \label{sec:distribute}}

\begin{figure}[!htbp]
	\centering
	\includegraphics[width=0.5\textwidth]{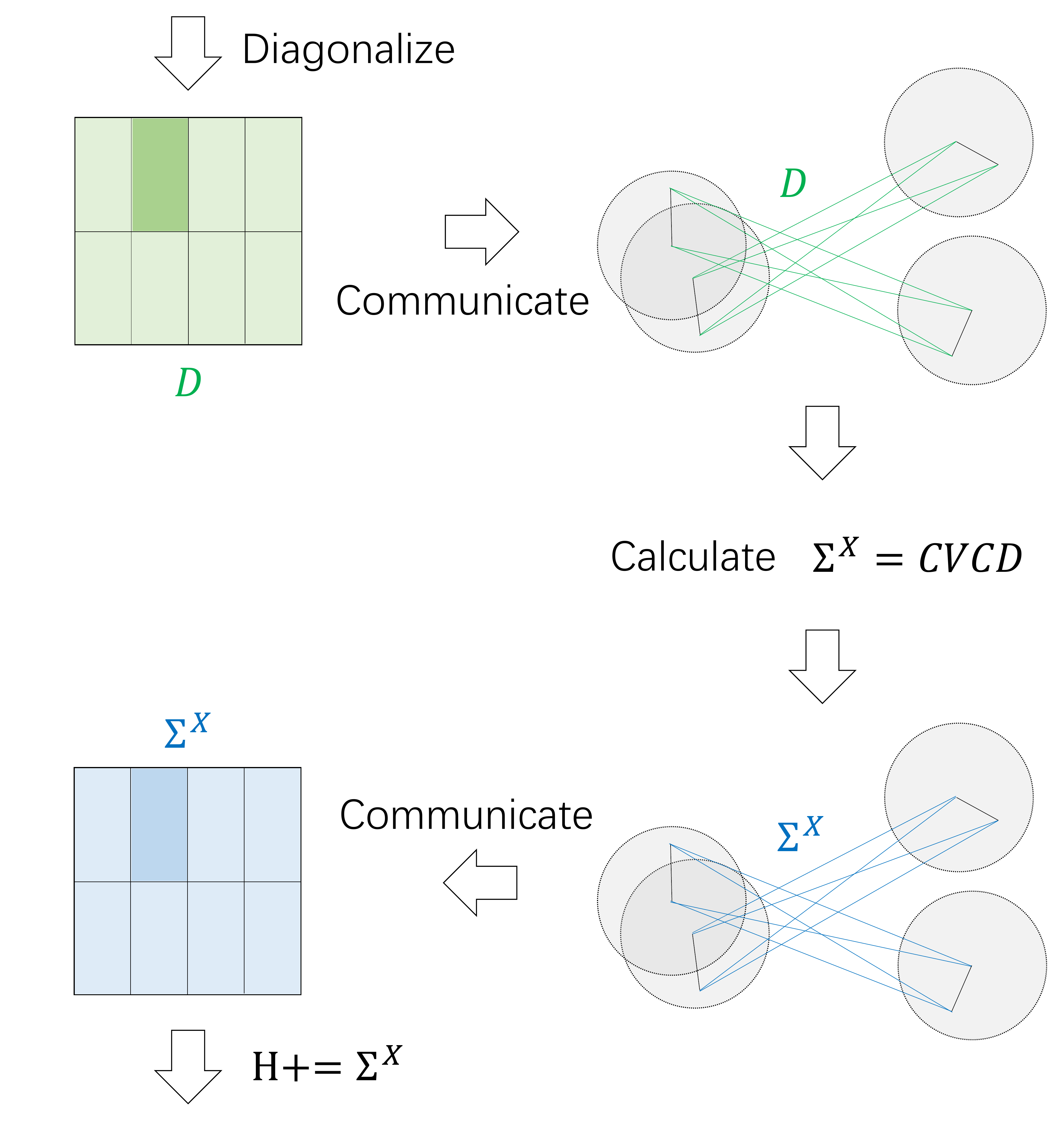}
	\caption{
	    Sketch of the communication and data redistribution of the density matrix $D$ and HFX matrix $\H$.
	    The left-hand side indicates the 2D cyclic-block form of the these two matrices, and the right-hand side indicates
	    their local atom-pair based distribution form.	}

	\label{fig:scheduling}
\end{figure}

According to the algorithms described in Sec.~\ref{sec:Hexx}, in particular Algorithm~\ref{lst:Hexx_4_AB},
our central parallelization strategy is to distribute the atom pairs \pair{A}{B} over different CPU cores.
For each \pair{A}{B} pair, we search the neighbouring atoms $F$ and $G$ within certain cut-off radii,
and calculate the corresponding $V_{A\alpha,B\beta}$, $C_{Aa,Ff}^{A\alpha}$ and $C_{Bb,Gg}^{B\beta}$ matrices.
For a given atomic structure, these matrices are calculated once and stored in memory beforehand,
since they depend only on the atomic structure and don't change during the self-consistent-field (SCF) cycles.
The density matrices $D_{Kk,Ll}$ and HFX matrices $\H_{Ii,Jj}$, on the other hand,
are updated at each iteration of the SCF loops.
The full density matrix $D$ is calculated after
diagonalizing the total Hamiltonian matrix $H$ in a 2D cyclic-block form, and hence initially also stored
in the same distributed form as $H$.
It is then redistributed over the \pair{A}{B} pairs via MPI communication tools\cite{MPI}, so that the needed matrix elements of
$D$ are locally available when building the HFX via Algorithm~\ref{lst:Hexx_4_AB}.
Once we have the needed
$C_{Aa,Ff}^{A\alpha}$, $V_{A\alpha,B\beta}$, $C_{Bb,Gg}^{B\beta}$, and $D_{Kk,Ll}$ 
matrices ready in each individual MPI process, the $\Ht$ matrices can be calculated independently
without any communication and the desired $\H_{Ii,Jj}$ matrix can be obtained from $\Ht$ matrices
(cf. Algorithm~\ref{lst:Hexx_4_AB}) via only light communications.
After the locally distributed $\H$ (based on atomic pairs) is calculated,
it will be transferred to the 2D cyclic-block form and added to the total Hamiltonian $H$.
The major communication processes are illustrated in Fig.~\ref{fig:scheduling}. The key feature of this
parallelization algorithm is that only the relatively cheap $\H$ and $D$ matrices need to be redistributed and communicated
among MPI processes, whereas the more expensive $C$ and $V$ matrices are evenly distributed over the MPI tasks and
no data communications for these are needed.

In principle, this part of calculations can be parallelized up to $M \times N_{\rm at}$ CPU cores,
where $N_{\rm at}$ is the number of atoms in a single unit cell,
and $M$ is the number of neighbouring atoms determined by the range of the $V$ matrix.
The load balancing totally depends on the distribution of \pair{A}{B} pairs.
A random distribution of the \pair{A}{B} pairs may cause severe load imbalance for wall time and computational bottleneck 
of memories, in particular for the non-uniform systems
where the number of neighbouring atoms may vary greatly for different atoms.
To overcome this difficulty, we developed two distribution schemes to improve the parallel efficiency.
The first one is called ``Machine-scheduling distribution'',
which is to improve the load balance based on computational time,
and the second algorithm is called ``K-means distribution'',
which can be used to reduce the memory usage. Depending on the availability of the
computational resources, we choose one or the other algorithm to distribute the computational load to achieve optimal performance.

The parallelization of HFX calculations in \textsc{ABACUS} is a hybrid mode based on processes and threads.
The MPI \cite{MPI} is adopted for parallelization over processes,
while OpenMP\cite{OpenMP} is
employed for parallelization over threads.
The \pair{A}{B} pairs are distributed to processes according to distribution schemes discussed below,
and then each process forks threads to claim and complete the distributed \pair{A}{B} pairs with 
\textit{dynamic schedule} in OpenMP for load balance.

\subsubsection{Machine-scheduling distribution \label{sec:scheduling}}

\begin{figure}[!htbp]
	\centering
	\includegraphics[width=0.5\textwidth]{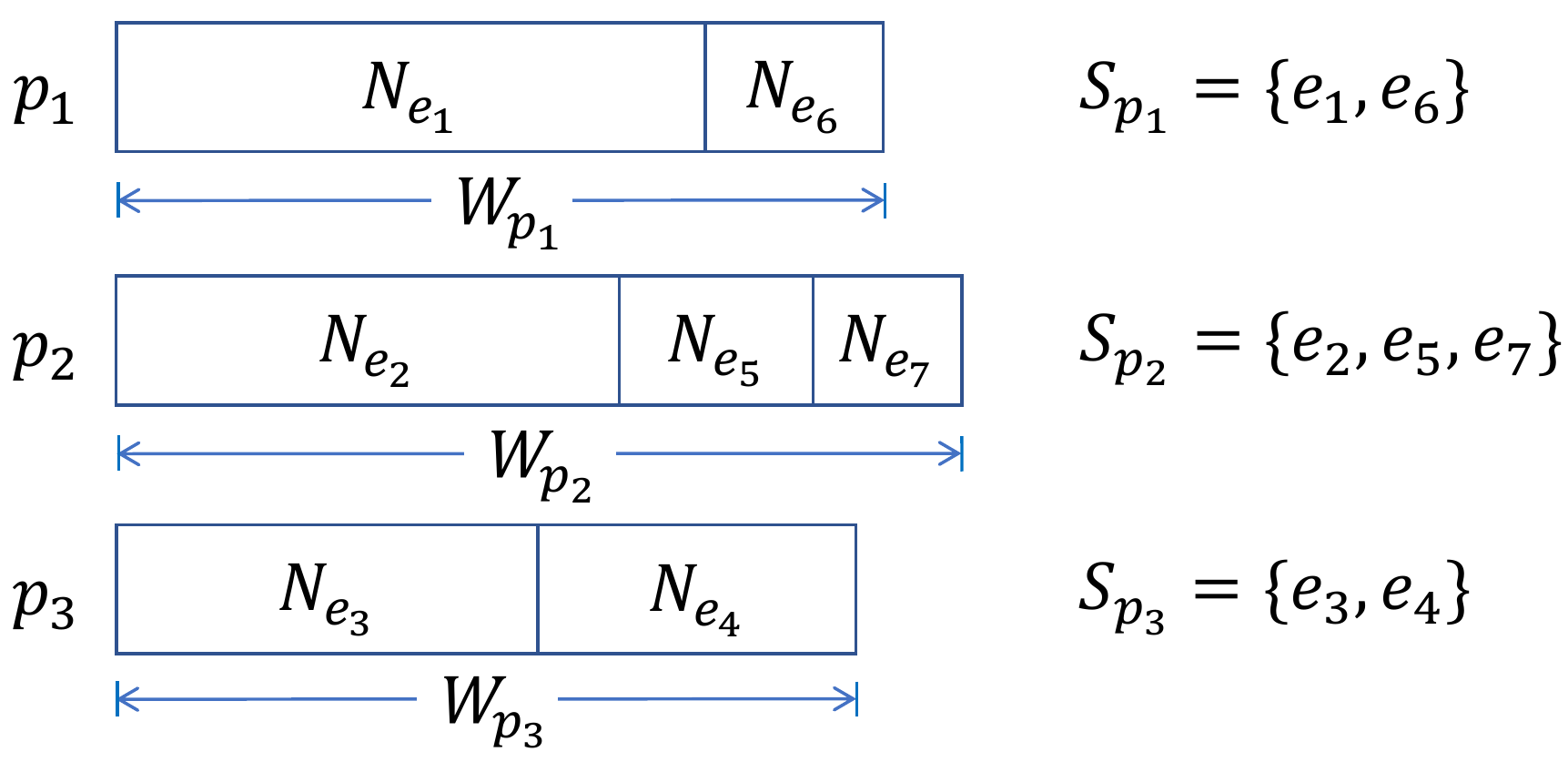}
	\caption{Graphical illustration of the process $p$, the list of tasks (``edges") $S_p$ on process $p$,
	       the weight of a give task (``edge $e$") $N_e$, and the total weight of tasks $W_p$ on process $p$.}
						 
	\label{fig:machine_scheduling}
\end{figure}

For each atomic pairs \pair{A}{B},
the size of the corresponding $\Ht$ ($ \sim CVCD$) matrix is proportional to $|\Neighbour{A}|*|\Neighbour{B}|$,
where $|\Neighbour{A}|$, $|\Neighbour{B}|$ are the numbers of neighbouring atoms of atom $A$ and $B$ respectively.
We can think of the atoms in the system as the vertices of an undirected graph,
whereas the atom pairs can be viewed as the edges.
The computational cost of each edge $e=\pair{A}{B}$ is roughly $N_e = |\Neighbour{A}|*|\Neighbour{B}|$,
which is the weight of the edge.
In this case, we try to distribute the {\it weighted} edges as evenly as possible among the CPU cores,
to minimize the maximum computational load on one core.
This becomes a classical machine scheduling problem\cite{kan1976machine}.
It's known that the global optimum solution of machine scheduling problem is NP-hard\cite{lenstra1977complexity},
and thus it is not possible to find the global optimum solution for large systems.
However, one can find an approximate solution by the greedy algorithm\cite{graham1966bounds,graham1969bounds}.
It can be proven that the maximum computational cost on one core given by the approximate solution
is not more than twice as that of the optimal solution. 
In practice, we find the greedy algorithm as presented in Algorithm~\ref{lst:scheduling} works very well.
In Algorithm~\ref{lst:scheduling}, $S_p$ denotes the list of tasks (``edges" in this case) in process $p$,
whereas $W_p$ denote the computational loads (``weight") on process $p$. The meanings and relationships between
$N_e$, $S_p$, $W_p$ are further graphically illustrated in Fig.~\ref{fig:machine_scheduling}.
\begin{algorithm*}
	\caption{The greedy algorithm of Machine-scheduling distribution}
	\label{lst:scheduling}
	\begin{algorithmic}[1]
		\Function{Machine-scheduling distribution}{}
			\ForAll{process $p$}
			\Comment{initialization}
				\State list of tasks $S_p = \emptyset$
				\State task load $W_p = 0$
			\EndFor

			\State sort $\{e\}$ in descending order of $N_e$
			\Comment{reduce unbalance}

			\ForAll{task $e$}
			\Comment{greedy algorithm}
				\State $p' = \argmin_p W_p$
				\State $S_{p'} = S_{p'} \bigcup \{e\}$
				\State $W_{p'} = W_{p'} + N_{e}$
			\EndFor
		\EndFunction
	\end{algorithmic}
\end{algorithm*}

\subsubsection{K-means distribution}
\label{sec:kmeans_method}

The ``Machine-scheduling distribution'' algorithm discussed above is suitable for achieving load balance in computation
time. However, it is not memory friendly. To understand this point, we first briefly analyze the memory consumption of 
the major arrays in parallel computation of the HFX matrix. 
The variation of the memory consumption upon increasing the number of compute nodes is a key factor that
affects the scalability of the calculation. The sparsity of $C$ in large systems is guaranteed by the locality of NAOs, and 
its memory consumption increases linearly with system size. In our implementation, the memory footprint of $C$ in each compute 
node can be reduced by increasing the number of compute nodes.
As for $V$, although it is dense for long-range Coulomb potential, the indices of the $V$ matrix are precisely those
used for parallel distribution. Hence the memory storage of $V$ in all processes is non duplicated,
and it memory consumption in each process is inversely proportional to the number of processes.
On the other hand, the parallel distribution of $D$ and $\H$ is much more involved,
and the memory consumption of these arrays may become the bottleneck for large size systems in parallel calculations.
In practice $\H$ consumes more memory than $D$, and hence we take $\H$ as an example to analyze the problem.  The 
the analysis of $\H$ applies to $D$ as well. 

\begin{figure}[!htbp]
	\centering
	\includegraphics[width=0.5\textwidth]{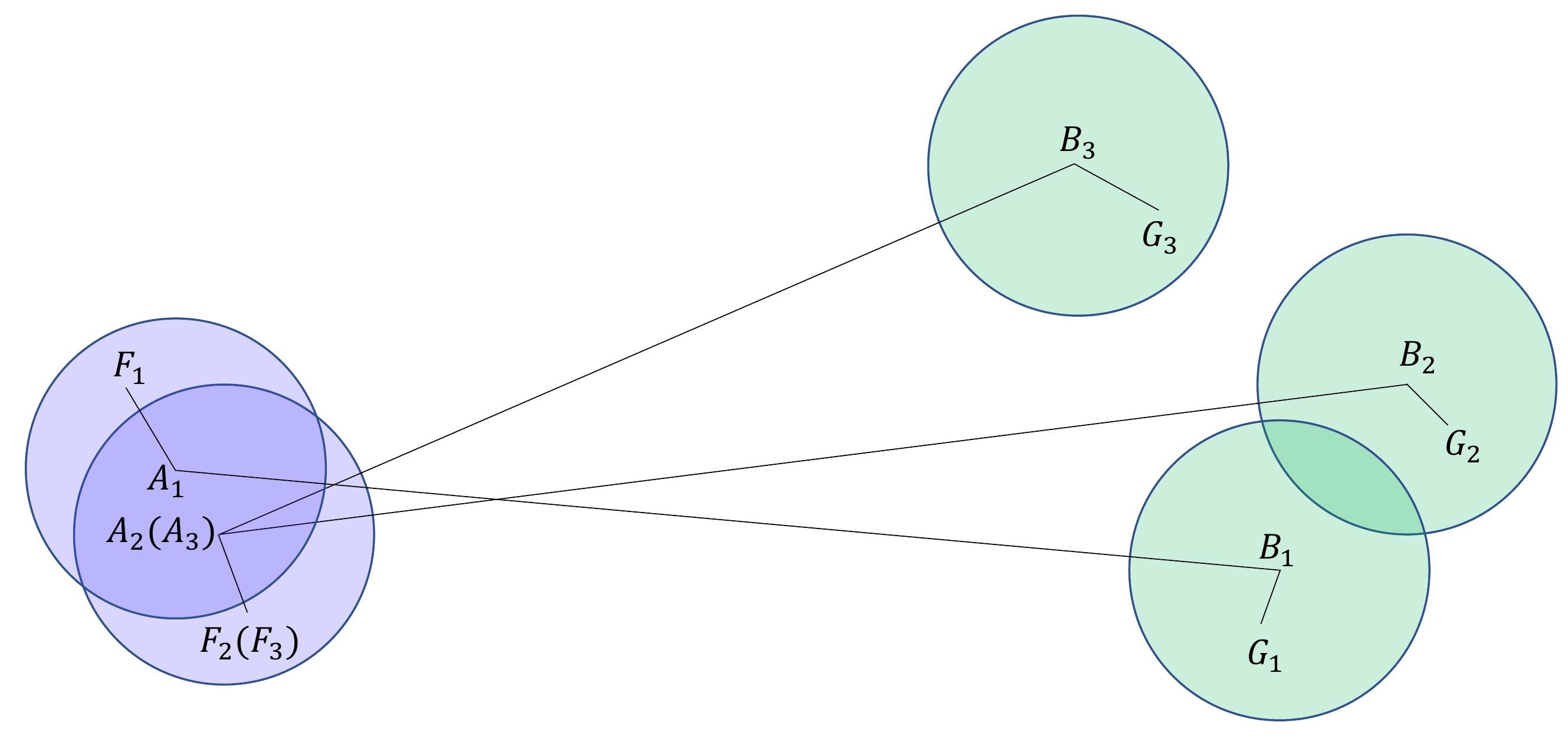}
	\caption{
	    Illustration of the memory consumption of $\H_{Ff,Gg}$ on one process. For an atom pair \pair{A}{B}, the set of $\H_{Ff,Gg}$ covers $\forall F \in \Neighbour{A}$ and $\forall G \in \Neighbour{B}$.
	    For a group of atom pairs $\{\pair{A}{B}\}$ distributed to one process, the set of $\H_{Ff,Gg}$ is the union of 
	    all $F,G$ atoms in the neighborhood of all these $\{\pair{A}{B}\}$ atom pairs.
	}
	\label{fig:kmeans}
\end{figure}

As illustrated in Algorithm~\ref{lst:Hexx_4_AB}, our actual implementation is based on the loop structure over 
atomic pairs \pair{A}{B}. For each pair \pair{A}{B}, one needs to evaluate contributions to four blocks of the
HFX matrices, i.e., $\H_{Ff,Gg}$, $\H_{Ff,Bb}$, $\H_{Aa,Gg}$ and $\H_{Aa,Bb}$, among which $\H_{Ff,Gg}$ is most
memory intensive because of the presence of different $F,G$ atoms in the neighborhood of \pair{A}{B} pair.
A pictorial illustration of the situation is presented in Fig.~\ref{fig:kmeans}. 
When a group of atom pairs \pair{A}{B} are distributed to one process $p$,
the set of $\H_{Ff,Gg}$ that need to be stored in process $p$ is given by
$\bigcupl_{\pair{A}{B} \in p} \bigcupl_{F\in\Neighbour{A}} \bigcupl_{G\in\Neighbour{B}} \H_{FG}$.
This means that the memory consumption of $\H_{Ff,Gg}$ on a process $p$ is proportional to the union of the neighborhood
regions of the atom $A$'s and $B$'s assigned to $p$, as illustrated in Fig.~\ref{fig:kmeans}. To minimize the memory consumption of $\H_{Ff,Gg}$, the number of different $F$ ($G$) atoms in the neighborhood of $A$ ($B$) atoms distributed on the same process 
should be as few as possible.
The ``Machine-scheduling distribution'' scheme doesn't consider the atom positions,
and the \pair{A}{B} pairs and their neighbors in one process may spread all over the supercell.
If this happens, simply increasing the number of compute nodes does not necessarily lead to the reduction
of the memory consumption of $H$ on one process.

Based on the above analysis, it is obvious that  minimizing the memory consumption for $\H_{Ff,Gg}$ on each process
amounts to minimizing $\bigcupl_A \Neighbour{A}$,
or equivalently, by maximizing the overlap of all $\Neighbour{A}$'s present in one process.
Considering that the atomic cut-off radii of all chemical elements in the actual calculations are very close if not equal,
this is essentially equivalent to making all $A$'s allocated to each process as close as possible.
The same principle applies to atom $B$.

Requiring the atom $A$'s ($B$'s) allocated to the same process as close as possible is a typical clustering problem --
an unsupervised learning problem in machine learning.
Specifically, we need to sort the atoms into several groups, in the three-dimensional Euclidean space, 
with the aims that the atoms in each group are as close as possible.
The K-means algorithm\cite{macqueen1967some} given in Algorithm~\ref{lst:kmeans} is chosen here to cluster atoms.

For short-range potential, such as HSE potential, etc.,
only \pair{A}{B} within certain ranges are needed,
whereas for long-range potential, such as HF etc.,
almost all pairs of \pair{A}{B} are needed.

\begin{algorithm*}
	\caption{The algorithm of K-means distribution}
	\label{lst:kmeans}
	\begin{algorithmic}[1]
		\Function{K-means distribution}{}
			\State $\tau_A$: coordinate of atom $A$
			\State $x_A$: category of atom $A$
			\State $\tau_x$: center coordinate of category $x$
			\While{unconverged}
				\ForAll{atom $A$}
					\State category $x_A = \argmin_x \{|\tau_A-\tau_x|\}$
				\EndFor
				\ForAll{category $x$}
					\State center coordinate $\tau_x = \text{average} \{\tau_A | x_A=x\}$
				\EndFor
			\EndWhile
		\EndFunction
	\end{algorithmic}
\end{algorithm*}

\section{Results and discussions}
In this section, we benchmark the efficiency
and scalability of our algorithm and implementations as discussed in Secs.~\ref{sec:methods} and \ref{sec:imple} 
via specific examples. We first demonstrate the effect of our screening techniques, which is the central
part of our implementation (Sec~\ref{sec:results_screening}). This is followed by an examination of the efficacy of the
two parallel distribution schemes (Sec~\ref{sec:results_parallel}).
A comprehensive benchmark study of the overall performance of our HFX implementation with respect to the system size and 
the number of CPU cores is presented in Sec.~\ref{sec:overall_perf}. Finally the advantage of our implementation for 
band structure calculations are discussed in Sec.~\ref{sec:results_band}. Note that we are concentrating here the
efficiency aspect of our implementation, and the accuracy aspect has been reported in 
a previous publication~\cite{Lin/Ren/He:2020}.

\subsection{Effects of screening\label{sec:results_screening}}

\subsubsection{Pre-screening of individual $C$, $V$ and $D$ matrices \label{sec:result_sparsity}}

We perform test calculations on a unit cell of $N_{\rm at}$ atoms with $N_{kx} \times N_{ky} \times N_{kz}$ $\k$-points,
which corresponds to a supercell of $N_{\rm at}\times N_{kx}\times N_{ky} \times N_{kz}$ atoms
with a single $\k$-point. In this setting, the atom $A$ can be restricted within the central unit cell,
whereas the atoms $B$, $F$, $G$ run over the entire supercell.
Thus,  all the $C$, $V$, and $D$ matrices have the same real-space data structure and hence the same sparsity 
as if we are dealing with a $\Gamma$-only supercell with $N_{\rm at}\times N_{kx}\times N_{ky} \times N_{kz}$ atoms. 
Therefore, a small unit cell with a dense $k$-point mesh is well suitable for testing
the effect brought about by screening out the insignificant elements of $C$, $V$ and $D$ matrices. Yet, 
compared to directly dealing with a large supercell, the computational cost can be 
significantly reduced, thanks to the translational symmetry.

As a specific test example, we performed HSE06\cite{krukau2006influence} calculations (with the screening parameter $\mu=0.11$) 
for Si crystal.
The lattice constant is chosen to be 10.236 Bohr, and a 8$\times$8$\times$8 $\k$-point mesh is used for Brillouin zone (BZ) integration.
The energy cut-off for determining the uniform real-space integration grid for charge density is set to be 240 Ry.
We use a NAO DZP basis [2$s$2$p$1$d$] for the one-electron basis set; for ABFs, an optimized [5$s$4$p$3$d$] set of orbitals \cite{Lin/Ren/He:2020} is used.
The cut-off radii of both NAOs and ABFs are set to 8 Bohr.

\begin{figure}[!htbp]
	\centering
	\includegraphics[width=\textwidth]{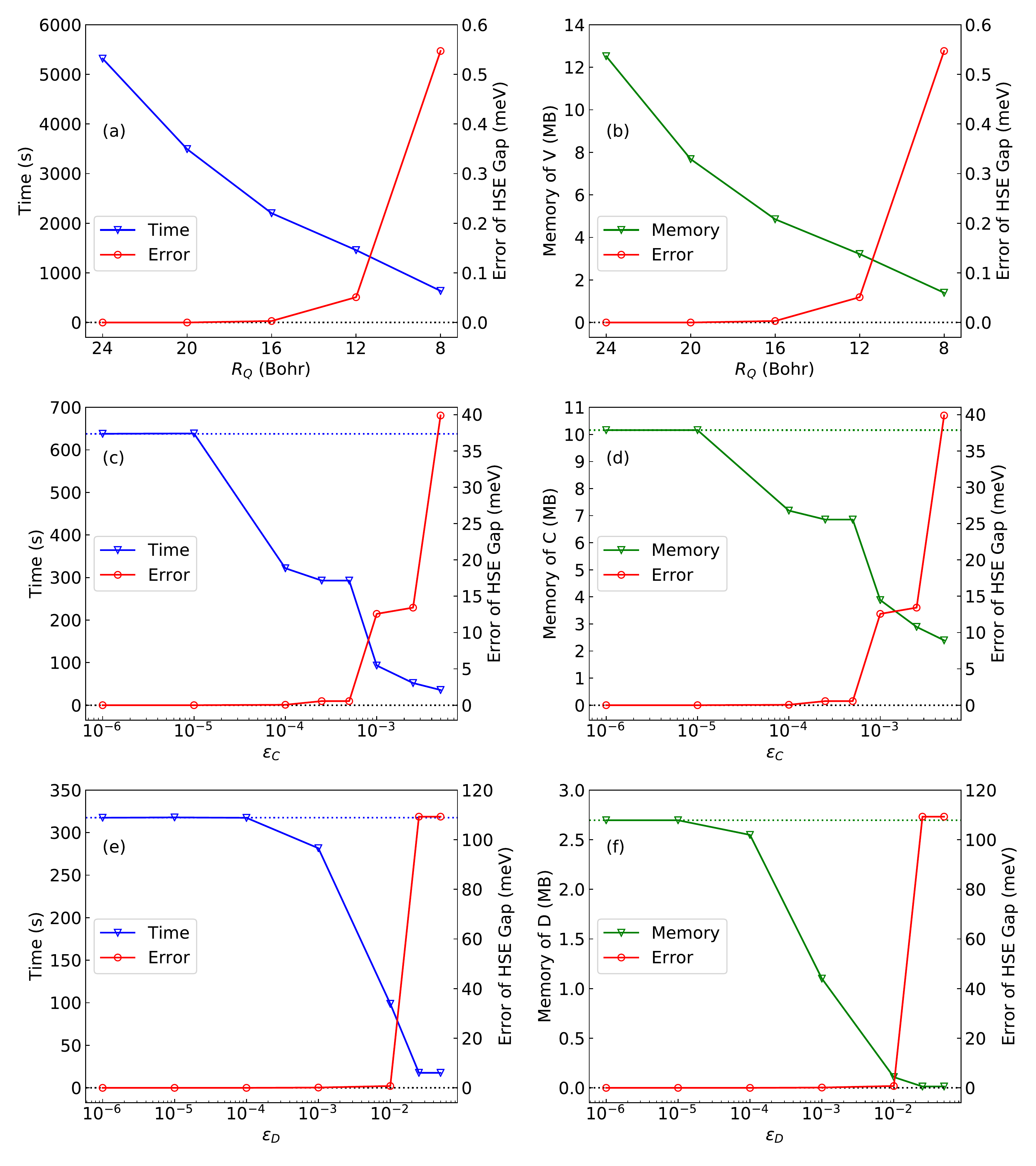}
	\caption{
	    Computation times (left panels) and memory footprints (right panels) as a function of the cut-off/thresholding parameters
	    in pre-screening the individual $V$ (upper panels), $C$ (middle panels), and $D$ (bottom panels) matrix. 
	    In all panels, the accompanying band gap errors are plotted
		as an indicator of the numerical accuracy of the calculations.}
	\label{fig:sparsity_CVD}
\end{figure}


Now we first check what happens if we pre-screen the small matrix elements of individual $C$, $V$, and $D$ matrices, i.e., 
the influence of the pre-screening on the numerical accuracy, the computation time, and the memory consumption. 
Here, the numerical accuracy of the calculations is
measured by the error of the obtained band gap, given by $\delta E_g = |E_g^\text{scr} - E_g^\text{ref}|$ where $E_g^\text{scr}$
is the bang gap value obtained with screening and  $E_g^\text{ref}$ is the reference value obtained without applying screening -- by setting
the thresholding parameters to zero, or in case of the $V$ matrix, by setting $R_Q$ to a big value.
Here the error of the band gap is used as an measurement of the computational accuracy, as it is more sensitive to
screening thresholds, compared to the variational quantities such as the ground-state total energy.
Fig.~\ref{fig:sparsity_CVD} presents the band gap errors, memory consumption for $C$, $V$, $D$ matrices, and the computation
time of evaluating the HFX matrix in one SCF iteration as a function of their respective screening thresholds or cut-off parameters. 
The calculation was performed on one Intel(R) Xeon(R) CPU (E5-2640 v2 @ 2.00GHz) core. As discussed in Sec.~\ref{sec:sparsity},
$\varepsilon_C$ and $\varepsilon_D$ are used to directly filter the subblocks of $C$ and $D$ matrices,
and a smaller values of $\varepsilon_C$ and $\varepsilon_D$ means more accurate calculations.
The pre-screening of the $V$ matrix, on the other hand, is controlled by a cut-off radius $R_Q$ of the Coulomb potential $Q(r)$ associated with the ABFs, as introduced in Sec.~\ref{sec:coulomb_matrix_screening}.
Obviously, a larger $R_Q$ corresponds to smaller numerical errors.

The test calculations are done in a successive way. Namely, we first examine the influence of $R_Q$, while setting $\varepsilon_C$
and $\varepsilon_D$ to be zero. Then, with a fixed value of $R_Q$ and zero $\varepsilon_D$ value, 
we look at the impact of $\varepsilon_C$. Finally,
with fixed $R_Q$ and a finite $\varepsilon_C$ threshold, we check the effect of $\varepsilon_D$.
Inspection of Fig.~\ref{fig:sparsity_CVD} reveals that, for all three matrices, there exists a window of parameter values,
within which there is almost no loss of accuracy (i.e., the obtained band gap staying essentially unchanged),
yet the computation time and memory consumption is significantly reduced.
In the HSE06 functional, the $Q(\bfr)$ itself is short-ranged, which allows for introducing a small cut-off radius 
for $Q(\bfr)$ to screen out the insignificant elements of the $V$ matrix.
As shown in Fig.~\ref{fig:sparsity_CVD}(a) and (b), when the cut-off radius $R_Q$ is reduced from 24 Bohr to 8 Bohr,
the incurred band gap error is negligible (less than 1 meV),
whereas the computation time is reduced from 5317 s to 637 s and the memory cost reduced from 12.5 MB to 1.4 MB.
Such a drastic reduction of the computation time and memory consumption certainly benefits from the short-ranged nature 
of the $V$ matrix in screened HDFs such as HSE.
In contrast, if large-range part of the Coulomb matrix is needed, such as in Hartree-Fock, or long-range-corrected HDFs, a
rather large $R_Q$ ($\sim 50$ Bohr) is needed, and
the memory storage of the $V$ matrix, as well as the computation time, will get significantly more expensive.


Fig.~\ref{fig:sparsity_CVD}(a) and (b) show the change of the computation time and the memory cost of the $C$ matrix
as a function of $\varepsilon_C$, with $R_Q=8$ Bohr and $\varepsilon_D=0$. 
It can be seen that, if $\varepsilon_C$ is set to 1$\times$10$^{-4}$, compared to $\varepsilon_C=0$ (no pre-screening)
the computing time is reduced from 637 s to 322 s (nearly $50\%$ reduction) and
the memory of the $C$ matrix from 10.2 MB to 7.2 MB (nearly $30\%$ reduction),
while the computed band gap is barely affected. Note that the zero elements in $C$ due to the finite cut-off radius of the AOs
have already been excluded at the very beginning the calculation, and the memory reduction recorded here
is due to the small but finite matrix elements of $C$. 
When increasing $\varepsilon_C$ from 5$\times$10$^{-4}$ to 2.5$\times$10$^{-3}$,
the computing time goes further down to 52 s (i.e., more than $90\%$ reduction) and
the memory cost of $C$ matrices down to 2.9 MB (i.e., more than $70\%$ reduction), but now
a visible band gap error of 13.4 meV is incurred.
However, further increasing $\varepsilon_C$ beyond 10$^{-3}$
leads to a rapid increase of the band gap error, which should be avoided.

Finally we check the influence of $\varepsilon_D$ with fixed $R_Q=8$ Bohr and $\varepsilon_C=10^{-4}$, and the obtained
results are reported in Fig.~\ref{fig:sparsity_CVD}(e) and (f).
One can see that, when increasing $\varepsilon_D$ from $0$ to $10^{-3}$,
the computation time is reduced from 322 s to 282 s, and the memory cost of $D$ is reduced from
2.7 MB to 1.1 MB. When further increasing $\varepsilon_D$ to $10^{-2}$,
the computation time is drastically reduced from 282 s to 99 s, and the memory consumption from 1.1 MB to 0.1 MB.
In the mean time, no noticeable change of the band gap is observed.
Such a significant saving in computation time and memory storage is enabled the fact 
that silicon crystal is an insulator, and its density matrix decays exponentially in real space.
However, if increasing $\varepsilon_D$ even further a little bit
(say, to 2.5$\times$10$^{-2}$), a rapid increase of the band gap error to $\sim$ 0.1 eV occurs.
Therefore, one needs to be very cautious when choosing the $\varepsilon_D$ parameter.
In practice, we found that a conservative value of $10^{-3}$ is safe and hence is recommended in
practical calculations. 




\subsubsection{Screening based on Cauchy-Schwarz inequalities}
\label{sec:result_cs}

After pre-screening individual $C$, $V$, and $D$ matrices, we can further apply the screening techniques based on Cauchy-Schwarz inequalities,
as discussed in Sec.~\ref{sec:schwarz} and Sec.~\ref{sec:cauchy}, to filter out those matrix elements that jointly lead to negligibly small $\Ht$ matrix 
elements. Any remaining insignificant elements of $C$, $V$, and $D$ matrices that passed the initial pre-screening step, will be identified and further excluded
here.

\begin{figure}[!htbp]
	\centering
	\includegraphics[width=\textwidth]{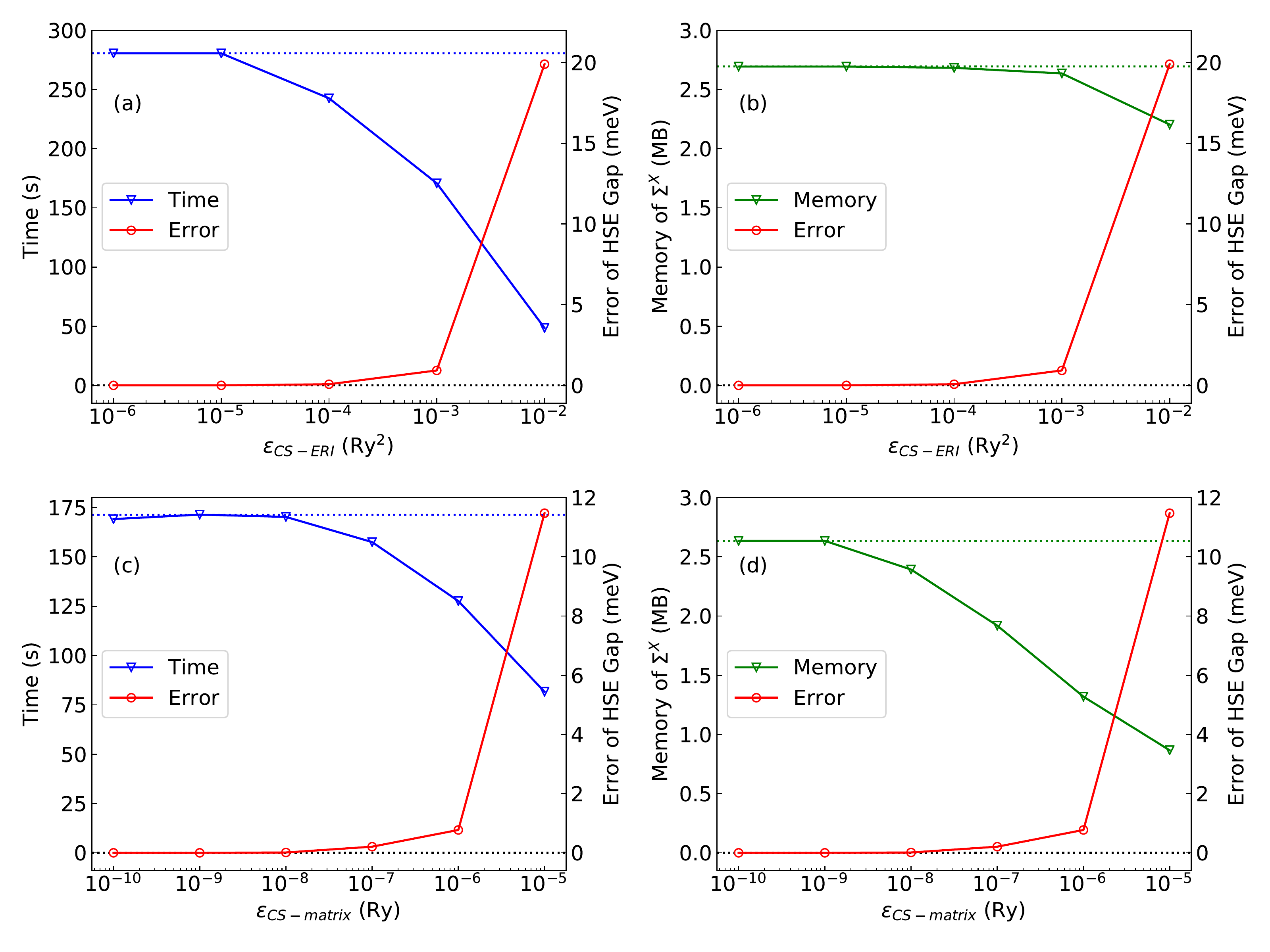}
	\caption{ Computation times (left panels) and memory footprints of $\H$ (right panels) as a function of the 
	     thresholding parameters of the ERI-based (upper panels) and ``matrix product" based (lower panels)
	     Cauchy-Schwarz screening. 
	    In all panels, the induced band gap errors are plotted
  		as an indicator of the numerical accuracy of the calculations. }
    	\label{fig:inequality}
\end{figure}

According to the screening workflow outlined in Algorithm~\ref{lst:exx_program}, we first apply the ERI-based Cauchy-Schwarz screening procedure and then
the ``matrix-product" based screening one. Note that, at this point, the pre-screening of individual $C$, $V$, and $D$ have already been 
performed with $R_Q=8$ Bohr, $\varepsilon_C=10^{-4}$, and $\varepsilon_D=10^{-3}$. Now, the memory storage of
the $\H$ matrix is used to measure the effect on the memory consumption due to the Cauchy-Schwarz inequality screenings. 
Fig.~\ref{fig:inequality}(a) and (b) present the computation time and the memory consumption of $\H$ as a function of
$\varepsilon_{\text{CS-ERI}}$--the thresholding parameter of the ERI-based Cauchy-Schwarz inequality ERI screening.
When setting $\varepsilon_{\text{CS-ERI}}=10^{-3}$, the induced error of the band gap is about 0.9 eV, while the computation time
is reduced from 282 s to 171 s. In this case, the influence on the memory cost of $\H$ is minor. Further increasing $\varepsilon_{\text{CS-ERI}}$
to $10^{-2}$ can accelerate the calculation by another a factor of 3, but the incurred error rises to about 20 meV, which is not recommended.

A final step is the Cauchy-Schwarz inequality matrix screening as outlined in Algorithm~\ref{lst:cauchy}. The computation time, 
the memory consumption of $\H$, and the band gap error as a function of the truncation threshold $\varepsilon_{\text{CS-matrix}}$ is shown 
in Fig.~\ref{fig:inequality}(c) and (d).
As $\varepsilon_{\text{CS-matrix}}$ increases from 0 to 10$^{-6}$, the computation time for evaluating HFX matrix is reduced from 172 s to 127 s
and the memory cost of $\H$ decreases from 2.6 MB to 1.3 MB.  In the meantime, the accompanying band gap error is only 0.77 meV.
Further increasing $\varepsilon_{\text{CS-matrix}}$ from 10$^{-6}$ to $10^{-5}$, the computation time is reduced to 82 s and memory of $\H$ to 0.9 MB,
but the band gap error is increased to 11 meV.

In summary, by applying both the pre-screening of individual matrices and the screening procedures based on Cauchy-Schwarz inequalities, the computation time
for evaluating the HFX matrix in one iteration is reduced by a factor of 40, whereas the total memory consumption of the four most memory intensive
matrices -- $C$, $V$, $D$, $\Sigma$  -- is reduced by a factor of 3. The accumulated error of the band gap, compared to the reference value without
applying any screening, is only 0.57 meV. Similarly, the incurred error in the absolute HSE06 total energy, which was not reported in the above analysis, 
is only 0.20 meV. (Note that the errors induced in individual steps might compensate each other, but are always in the same order of magnitude.) 
Obviously, if one can tolerate bigger errors, say 10 meV in band gap, the savings in computation time and memory storage will be even more significant.

\subsection{Parallel distribution schemes\label{sec:results_parallel}}

\subsubsection{Machine-scheduling distribution}

In parallel computing, keeping good load balance among the processes is a key requirement to enable massively parallel
calculations. In our implementation, this is achieved by
the so-called ``Machine-scheduling distribution'', as described in Sec.~\ref{sec:scheduling}. To
assess the performance of this distribution scheme, we compare it with a straightforward parallelization
scheme where the atom pairs are distributed randomly, only requiring that the numbers of atom pairs
assigned to each process are equal.

The test system chosen here is a DNA fragment containing 788 atoms, composed of 12 AT basis pairs, as studied in Ref.~\citenum{liu2019dft}.
The load balance is measured by the ratio between the maximal time consumed on one process ($\Delta t_\text{max}$) and the average time over all processes 
($\Delta t_\text{av}$). 
If the load balancing is perfect, $\Delta t_\text{max}/\Delta t_\text{av}$ ratio should equal 1; otherwise this ratio will be larger than 1. Obviously, 
the larger the ratio is, the worse the load balance.  We note that the global communication time between different processes is accounted for here, 
as it is the ``common" time shared by all processes. 

As a numerical experiment, we carried out six independent HSE06 calculations, respectively, on 1, 2, 4, 8, 16, and 32 compute nodes, each with 24 CPU cores. 
Consistent with the computer architecture, the calculations are parallelized using 1, 2, 4, 8, 16, and 32 MPI processes, each process consisting of
24 threads.
When running on one compute node (1 process $\times$ 24 threads), 
both ``Machine-scheduling distribution'' and ``random distribution'' schemes perform perfectly, with  $\Delta t_\text{max}/\Delta t_\text{av}$ ratio 
being essentially one, as it should be.
As the number of processes increases,
the $\Delta t_\text{max}/\Delta t_\text{av}$  ratio of the ``random distribution'' scheme grows gradually, reaching 2.2 when 32 nodes (768 CPU cores) are used,
meaning that the maximum computing time on one process is more than twice of the average computing time.
In contrast, the $\Delta t_\text{max}/\Delta t_\text{av}$ ratio of the ``Machine-scheduling distribution'' scheme stays very close to 1, increasing only slightly
when more CPU cores are used. With 32 processes (768 CPU cores), the wall time is still less than 1.2 times of the ideal time.
These results clearly demonstrate that, with the ``Machine-scheduling distribution'' scheme, one can achieve excellent load balancing in parallel computing
even for very inhomogeneous systems.

\begin{figure}[!htbp]
	\centering
	\includegraphics[width=0.5\textwidth]{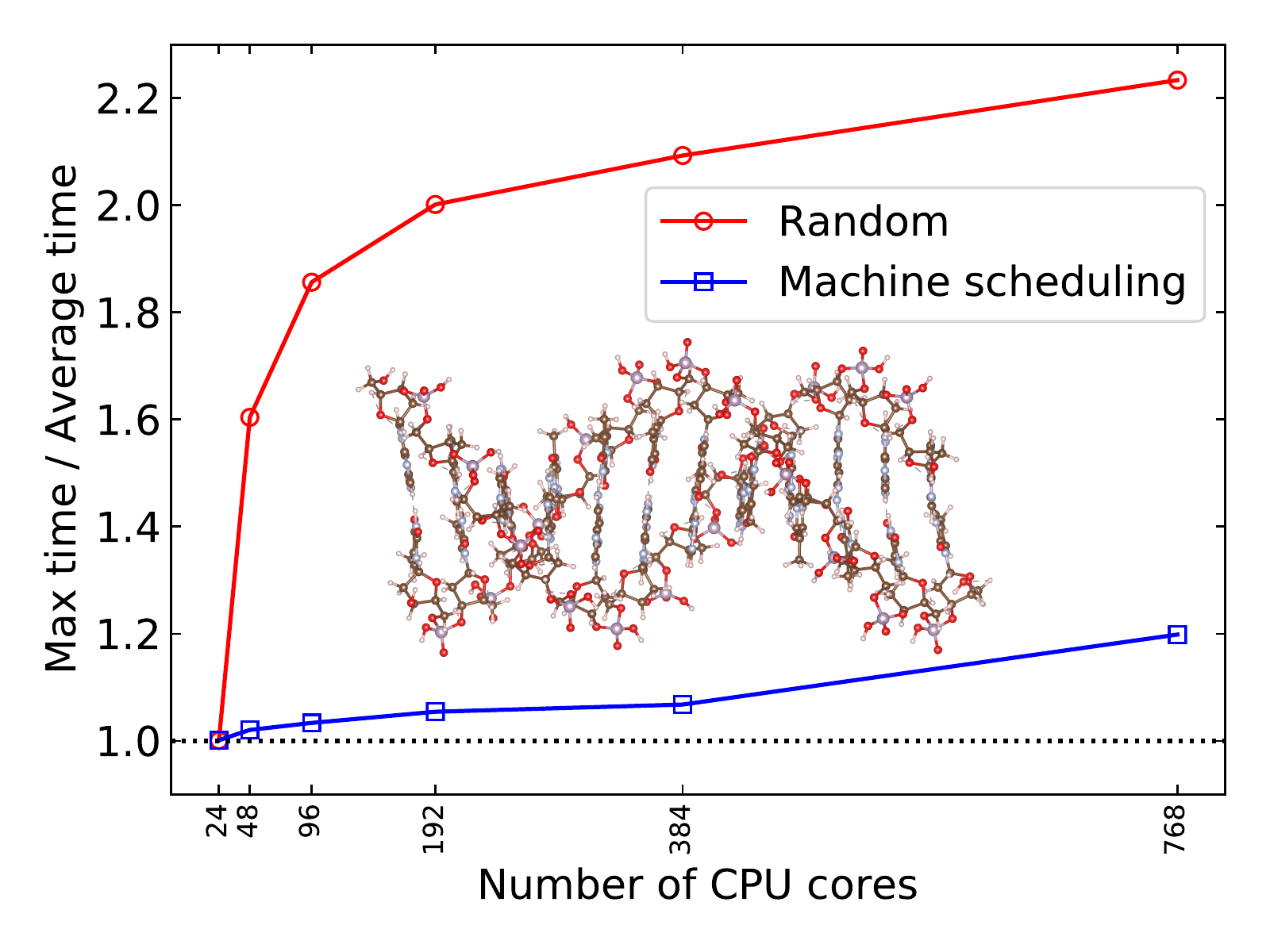}
	\caption{The ``Max time/Average time'' ($\Delta t_\text{max}/\Delta t_\text{av}$) ratio for the ``random distribution''
		(red circles) and ``Machine-scheduling distribution'' (blue squares) as a function of the number of CPU cores. 
		 Test calculations are done for a DNA fragment.
		}
	\label{fig:scheduling3}
\end{figure}

\subsubsection{K-means distribution \label{sec:result_kmeans}}

As discussed in Sec.~\ref{sec:kmeans_method}, in addition to the ``Machine-scheduling distribution" scheme intended to improve
the load balance of computation time, we offer an alternative, the so-called ``K-means distribution" scheme to reduce the memory footprints. 
In our implementation, the memory consumption scales roughly linearly 
with the size of the unit cell, and can become a bottleneck for very big supercells. In these cases, the ``K-means distribution" scheme 
enables calculations that would otherwise not run.

\begin{figure}[!htbp]
	\centering
	\includegraphics[width=0.5\textwidth]{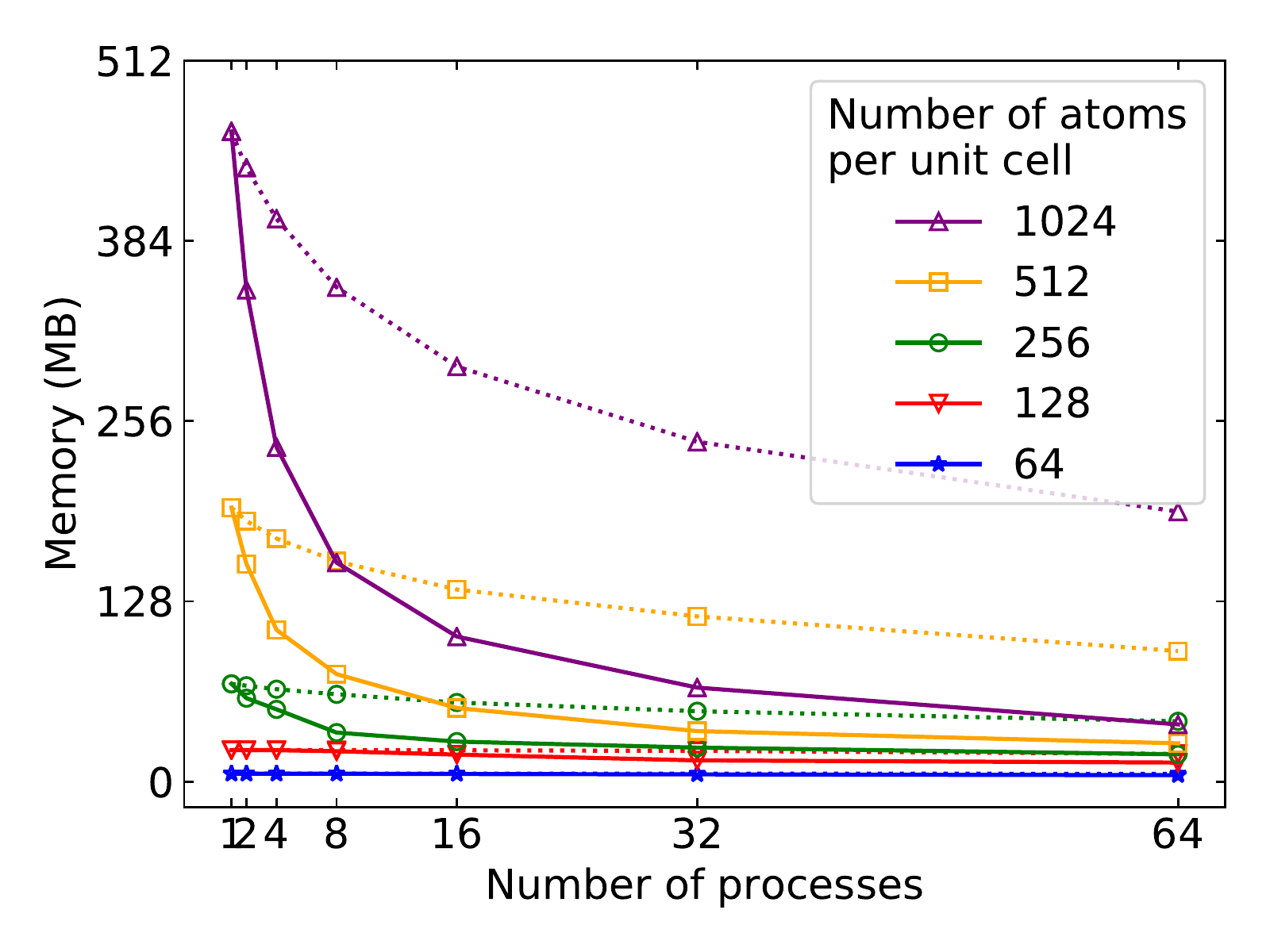}
	\caption{
		The maximal memory footprint of $\H$ among all runtime processes as a function of the number of MPI processes, 
		for different unit cell sizes.
		The solid lines correspond to  the ``K-means distribution'' scheme whereas the dashed lines represent
		the ``random distribution'' scheme.}
	\label{fig:kmeans_H}
\end{figure}

To test the performance of the ``K-means distribution" scheme, we carried out a series of $\Gamma$-only HSE06 calculations for Si crystal with different
unit cell sizes and increasing number of MPI processes.  
The maximal memory footprints for the $\H$ matrix per process for both ``K-means distribution'' and ``random distribution'' schemes
are presented in Fig.~\ref{fig:kmeans_H} as a function of the number of MPI process (again each process running on 24 CPU cores with shared memory)
Different curves in Fig.~\ref{fig:kmeans_H} correspond to different sizes of the unit cell, containing 64, 128, 256, 512, and 1024 Si atoms, respectively.
The computation parameters (basis sets, cut-off energy, and cut-off radii) are the same as those used in Sec.~\ref{sec:result_sparsity}.
The screening thresholds (or truncation parameters) are chosen to be 
$R_{Q}$=8 Bohr, $\varepsilon_{C}=10^{-3}$, $\varepsilon_{D}=10^{-2}$, $\varepsilon_{\text{CS-ERI}}=10^{-3}$,
$\varepsilon_{\text{CS-matrix}}=10^{-5}$.
Tests showed that the incurred band gap error with these screening parameter settings is below 10 meV.

As discussed in Sec.~\ref{sec:kmeans_method}, in our current parallelization algorithm, the same sublocks of the $\H$ matrix 
have to be stored in different processes, leading to a duplication of the memory storage of the $\H$ matrix. This may become
a bottleneck for large-scale calculations.
Fig.~\ref{fig:kmeans_H} clearly demonstrates the supremacy of the ``K-means distribution'' scheme (solid lines)
over the unoptimzed ``random distribution scheme"  (dashed lines) in reducing the memory cost as the number of processes 
increases. With two processes, the ``K-means distribution"
scheme gains a factor of 1.25 memory saving, while this number steadily increases to 4.7 for 64 processes. 
Thus, the ``K-means distribution" scheme can be invoked when there is a lack of memory, in particular in cases of massively 
parallel calculations.


\subsection{Overall performance \label{sec:overall_perf}}

\begin{figure}[!htbp]
	\centering
	\includegraphics[width=0.8\linewidth]{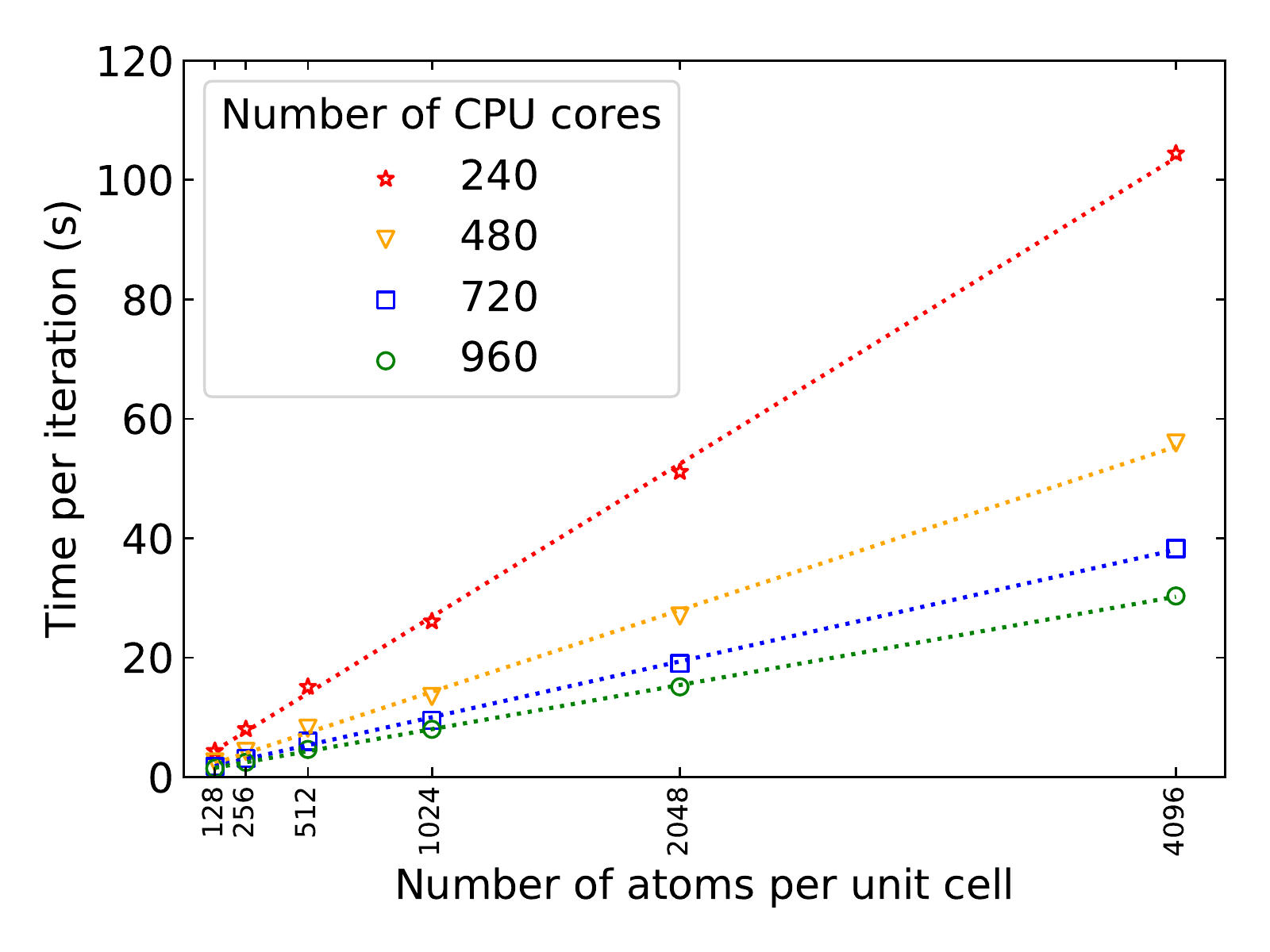}
	\caption{
		Computation time for building the HFX matrix per SCF iteration as a function of the system size for different
		amount of CPU resources. Test systems are Si crystal with various supercell sizes.}
	\label{fig:scaling_atom}
\end{figure}


To document the scaling behavior of the computational time for building the HFX matrix
with the increase of system size and computing resource,
we performed $\Gamma$-only HSE06 calculations for Si crystal with different supercell size and
using different number of CPU cores. The computation time recorded here is 
the wall time per iteration of evaluating and transmitting the $\H$ matrix,
which is the most time-consuming part of the entire HDF calculation for system sizes
tested so far (4906 atoms per unit cell).
The basic computation parameters and screening thresholds are identical to those used
in Sec.~\ref{sec:result_kmeans}. Since there is no memory shortage problem in these
calculations, the ``Machine-scheduling distribution'' scheme is used to achieve
good load balances.
All calculations are done on the Tianhe-2 supercomputer,
where each node  has two Intel(R) Xeon(R) CPUs (E5-2692 v2 @ 2.20GHz).

Fig.~\ref{fig:scaling_atom} presents the wall times
as a function of the system size, i.e., the number of atoms in  the supercell, for different numbers of CPU cores.
We have tested from the smallest system containing 128 Si atoms to the largest system containing 4096 Si atoms.
The size of the systems we have looked at ranges from 128 atoms up to 4096 atoms in the supercell.
The calculations employs 10, 20, 30 and 40 processes $\times$ 24 threads,
running on 240, 480, 720 and 960 CPU cores, respectively.
As can be seen from Fig.~\ref{fig:scaling_atom}, in all parallel runs with different processes,
the wall time scales almost linearly with the number of atoms in the calculations.
A linear fit of the data obtained using 960 CPU cores yields
$t$=0.0072$N_{\rm at}$+0.6401 (coefficient of determination $R^2$=0.9997).
Although the linear scaling behavior is expected from the underlying algorithms,
this benchmark test proves the efficacy of our implementation in \textsc{ABACUS}. 
Furthermore, the absolute timings presented in Fig.~\ref{fig:scaling_atom} indicate the prefactor of
our linear-scaling algorithm is rather small -- a feature that is vitally important for the usability
of the code for practical calculations.

\begin{figure}[tbp]
	\centering
	\includegraphics[width=0.8\linewidth]{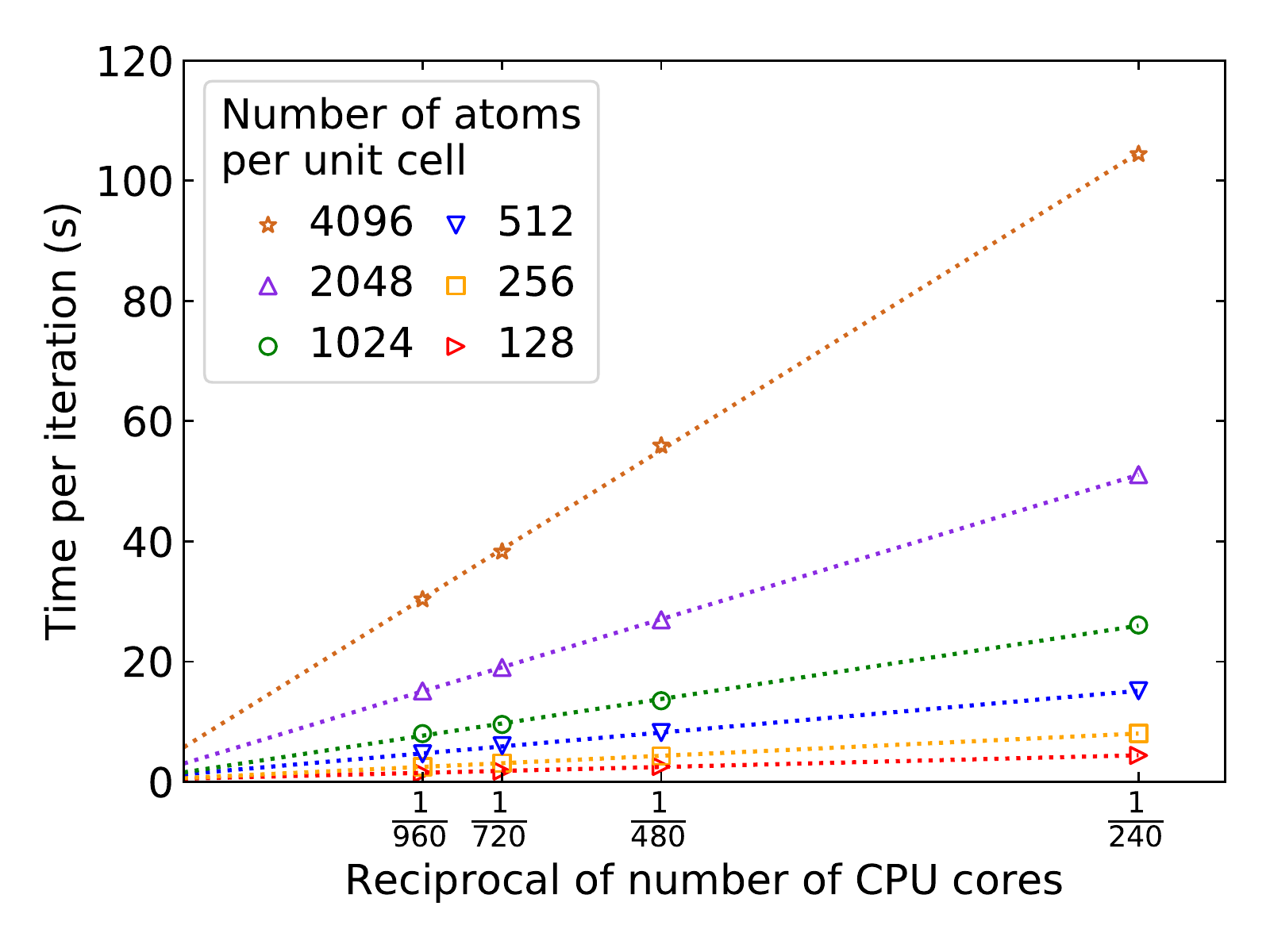}
	\caption{
		Computation time for building the HFX matrix per SCF iteration as a function of the number of CPU cores for
		different supercell sizes. Test systems are Si crystal with various supercell sizes.}
	\label{fig:scaling_core}
\end{figure}

In Fig.~\ref{fig:scaling_core}, the wall times of the calculations presented above are re-plotted as 
a function of (the inverse of) the number of CPU cores. 
It can be seen that, for all system sizes,
the computation time decreases linearly as the number of CPU cores increases.
Even for the smallest system containing only 128 Si atoms, the linear scaling is still perfect up to 960 CPU cores.
For the 4096-atom system -- the largest one tested in this work, the calculation takes about 104 s for one HFX evaluation
step using 240 CPU cores, and the computation time is reduced to about 30 s, if 960 CPU cores are used.
A linear fit for the data of the 4096-atom system yields
$t$ =23731 $N^{-1}_{\text{core}}$ + 5.7447 ($R^2$ = 0.9998).
The small prefactor in our linear-scaling implementation with respect to the system size and the
excellent parallel efficiency allows to tackle large-scale systems with relative ease. 
With our implementation in \textsc{ABACUS}, HSE06 calculations for systems with a few thousands of 
atoms can be routinely performed. In fact, given the linear-scaling behavior of our implementation
with respect to the system size and the number of CPU cores, there is in principle to handle systems
with tens of thousands of atoms. However, in that case, diagonalizing the Hamiltonian matrix will
become the new bottleneck -- a challenge that is common to all KS or gKS-DFT calculations.



\subsection{Band structure calculations\label{sec:results_band}}

One of the additional advantage of the present implementation is that the electronic band structures of HDF calculations
can be easily obtained. We briefly discuss this point in this subsection. In our implementation, we first obtain the
HFX matrix in real space -- $\H_{Ii,Jj}$, as indicated in eq~\eqref{eq:HFX_ERIs}, and then Fourier-transform it to $\bfk$ space. 
Note that in our notational system,
the atom $I$, and $J$ can be located in different unit cells $\bfR_I$ and $\bfR_J$, and thus can be rewritten as
\begin{equation}
    \H_{Ii,Jj} = \H_{\aR{I}(\bfR_I)i,\aR{J}(\bfR_J)j} = \H_{\aR{I}i,\aR{J}j}(\bfR_J-\bfR_I)=\H_{\aR{I}i,\aR{J}j}(\bfR)\,
\end{equation}
where $\aR{I}$, and $\aR{J}$ denote the atomic indices in one unit cell and $\bfR=\bfR_J-\bfR_I$. For most systems,
the exchange interactions are short-ranged, meaning that the matrix elements of $\H_{\aR{I}i,\aR{J}j}(\bfR)$ 
are vanishingly small for $|\bfR|>R_\text{max}$ where $R_\text{max}$ is certain critical length.


After self-consistent HDF calculations, we can obtain the real-space HFX matrix $\H_{\aR{I}i,\aR{J}j}(\bfR)$ for all lattice vectors
with $|\bfR|<=R_\text{max}$, and merge it with the local part of the gKS Hamiltonian to get the \textit{full} Hamiltonian. 
Once the the {\it full} gKS Hamiltonian $H_{\aR{I}i,\aR{J}j}(\bfR)$ in real-space is obtained, one can readily
construct the Hamiltonian at arbitrary $\bfk$ points in NAO basis sets,
\begin{equation}
	H_{\aR{I}i,\aR{J}j}(\k) = \suml_{|\bfR|<=R_\text{max}} e^{i\k\cdot \R} H_{\aR{I}i,\aR{J}j}(\R).
	\label{eq:Ham_rs2kspace}
\end{equation}
The reason that the lattice summation in eq~\eqref{eq:Ham_rs2kspace} can also be restricted below $R_\text{max}$ is
because the local part of the gKS Hamiltonian is
more short-ranged than the HFX part. Now, given that the Hamiltonian matrix at arbitrary $\bfk$ points is readily available 
from eq~\eqref{eq:Ham_rs2kspace}, the band energies along desired paths in $\bfk$ space can be obtained by a
one-shot diagonalization through non-SCF calculations. Compared to the plane-wave formalism,
the diagonalization at the final step is rather inexpensive, due to the much reduced basis size in the NAO framework.
Therefore, one does not need to invoke the band interpolation techniques here \cite{Pickett/Krakauer/Allen:1988,Shirley:1996b}, 
as is usually within the plane-wave approach. Such a real-space algorithm is also of great advantage if very dense $\bfk$ grids
are needed, e.g., when calculating the optical adsorption spectra.

\begin{figure}[tbp]
	\centering
	\includegraphics[width=0.45\linewidth]{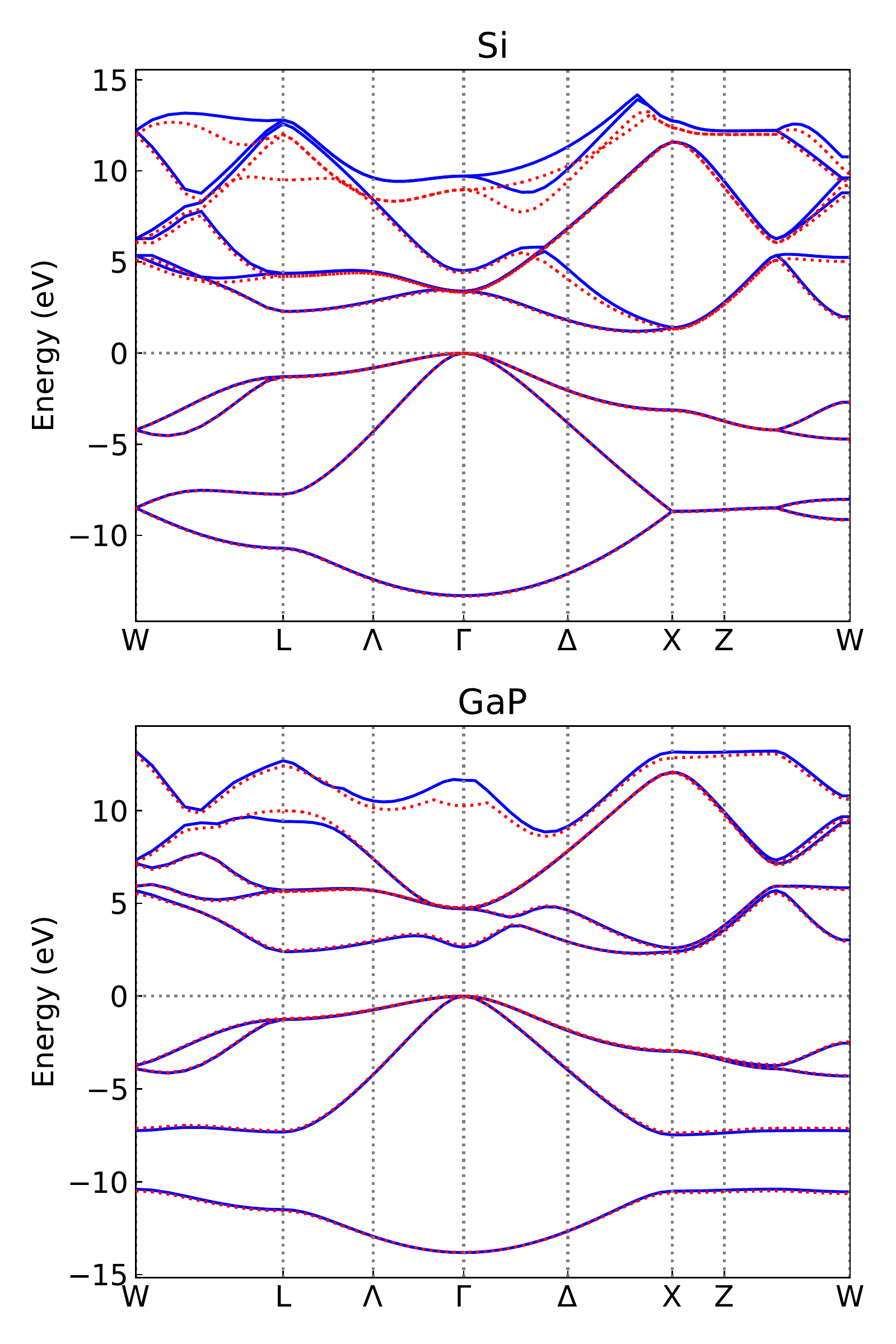}
	\caption{HSE06 electronic band structures of Si (upper panel, diamond structure) and GaP (lower panel, zinc blende structure) calculated 
	 using \textsc{ABACUS} (blue solid lines) and FHI-aims (red dashed lines) codes. The experimental lattice parameters are used
	 for both materials. A $8\times 8\times 8$ $\bfk$ grid is used in the BZ integration, and the NAO basis sets used by the two codes in
	 the present calculations can be found in the text.}
	\label{fig:band_str}
\end{figure}

To check the validity of our approach for HDF band structure calculations,
in Fig.~\ref{fig:band_str} we present the HSE06 band structures for Si and GaP crystals as obtained by \textsc{ABACUS},
in comparison with the corresponding FHI-aims\cite{Blum/etal:2009,Ren/etal:2012,Levchenko/etal:2015} results, which 
are taken as the reference here. The valence-only DZP basis sets ($2s2p1d$ for Si and P, and $2s2p2d1f$ for Ga) are used 
in \textsc{ABACUS} calculations, whereas the so-called ``tight" setting
is used in FHI-aims calculations, corresponding to all-electron $4s3p2d1f1g$ basis set for Si and P, and $5s4p2d1f$ basis set 
for Ga). Despite the different (pseudopotential versus all-electron) descriptions of core-valence interactions and 
different basis sizes, the valence and low-lying conduction bands (and hence the band gap) obtained using the two codes 
agree with each other rather well. The remaining discrepancy for the high-lying conduction bands is expected due to the 
relatively smaller basis size used in \textsc{ABACUS} calculations. 
The agreement will get further improved if one employs the TZDP basis sets in \textsc{ABACUS} calculations. 
A more comprehensive comparison study of the HSE06 band gaps obtained using different computer codes, as well as the
influence of the basis sets, can be found in Ref.~\cite{Lin/Ren/He:2020}.

\section{Summary}

In summary, we presented an efficient, linear-scaling implementation
for building the (screened) HFX matrix for periodic systems within the framework of
NAO basis functions. The implementation was based on the LRI approximation, coupled with our own procedures for
constructing the ABFs. The numerical accuracy of such an approximation for periodic HDF calculations has been
systematically benchmarked in Ref.~\citenum{Lin/Ren/He:2020}.

In this work, we described the numerical details behind our implementation, in particular how we choose
the loop structure over the atom pairs, and how we exploit the sparsity of the key matrices, including expansion coefficients, the 
(screened) Coulomb matrix, and the density matrix. In the latter case, a multi-level screening procedure is
employed which ensures the insignificant elements of these matrices can be efficiently screened out as much as possible,
leading to a linear-scaling build of the HFX matrix with a rather small prefactor. We discussed two parallel distribution schemes,
which can be invoked to achieve the best load balance for computation time, or alternatively, to reduce the memory consumption when
there is a shortage of memory. Benchmark calculations for Si crystal with supercells up to 4096 atoms confirms the linear-scaling behavior 
of the computation cost with respect to the system size, whereas calculations with increasing computing resources demonstrate the excellent
parallel efficiency up to $10^3$ CPU cores. 

Our implementation was carried out in \textsc{ABACUS}, but the same techniques can be easily utilized by other local-orbital based computer code package.  
With our present implementation in \textsc{ABACUS}, HDF calculations for systems with a few thousand atoms can be routinely done with modest computing
resources. We that expect our implementation will find important applications in disordered systems, defects, and heterostructures 
where large supercells are needed.

\begin{acknowledgement}
We thank Liu Xiaohui, Chen Junshi, Shen Yu and Shi Rong for helpful discussions.
The work is supported by the National Key Research and Development Program of China (Grant No. 2016YFB0201202)
and National Natural Science Foundation of China (Grant Numbers 11774327, 11874335).
The numerical calculations have been partly done in National Supercomputer Center in GuangZhou
and partly on the USTC HPC facilities.
\end{acknowledgement}

\appendix
\section{Appendix}

\subsection{Ordering of matrix multiplications to evaluate $\Ht$ matrices \label{sec:matrix_multiply}}
The key step in building the HFX matrix is to evaluate the \Ht matrices, formally introduced in
eq~\eqref{eq:HFX_individual_term}. There are four variants of them (cf. eq~\eqref{eq:HFX_LRI}), depending on where
the ABFs are located, but all can be seen as the contributions to $\H_{Ii,Jj}$ from the
the atom pair $\pair{K}{L}$. They are formally given by $\Ht\sim CVCD$, i.e., a sequence
of matrix products involving $C$, $V$, $D$ matrices. Obviously, the actual order of performing the matrix multiplication
is important here, since it will affect the overall computational cost. 

The issue can be analyzed for a given set of four atoms -- $I,J,K,L$. In this case, omitting the atomic
indices, the \Ht matrices are rank-2 tensors and can be calculated as
  \begin{equation}
     \Ht_{ij} = \suml_{kl} \suml_{\alpha\beta} C_{ik}^{\alpha} V_{\alpha\beta} C_{jl}^{\beta} D_{kl}
     \label{eq:matrix_multiplication}
  \end{equation}
For the convenience of analysis,
we assume that all atoms have the same number of AOs ($n_{\phi}$) and ABFs ($n_P$).
Here we emphasize that $n_{\phi}$ and $n_P$ refer to the number of basis functions per atom (and not
per unit cell). Therefore,  in eq~\eqref{eq:matrix_multiplication} $C$ is a $n_{P} \times n_{\phi} \times n_{\phi}$ 3rd-rank tensor,
and $V$ and $D$ are, respectively, $n_{P} \times n_{P}$ and $n_{\phi} \times n_{\phi}$ matrices.

Given the expression in eq~\eqref{eq:matrix_multiplication}, one may recognize that there are five different ways of 
ordering the matrix multiplications.
The computational costs associated with the five orderings are list in Table~\ref{tab:multiply},
among which the most efficient two are obviously $(CV)(CD)$ and $C((VC)D)$,
due to the fact that $n_{P}$ is several times larger than $n_{\phi}$. After considering
data structures adopted in the code, cache optimization,
and the use of BLAS\cite{lawson1977basic,dongarra1988extended,dongarra1990algorithm}
library, we finally chose $C((VC)D)$ as the matrix multiplication order in our implementation.

\begin{table}[!htbp]
	\centering
	\caption{computational cost of order of matrix multiplication}
	\label{tab:multiply}
	\begin{tabular}{cc}	\hline
		order of matrix multiplication	&	computational cost		\\\hline
		$(CV)(CD)$	&	$                       n_{P}^2 n_{\phi}^2 + 8 n_{P} n_{\phi}^3$	\\
		$C((VC)D)$	&	$                       n_{P}^2 n_{\phi}^2 + 8 n_{P} n_{\phi}^3$	\\
		$C(V(CD))$	&	$                     4 n_{P}^2 n_{\phi}^2 + 8 n_{P} n_{\phi}^3$	\\
		$((CV)C)D$	&	$  n_{P} n_{\phi}^4 +   n_{P}^2 n_{\phi}^2 + 4 n_{\phi}^4$		    \\
		$(C(VC))D$	&	$  n_{P} n_{\phi}^4 +   n_{P}^2 n_{\phi}^2 + 4 n_{\phi}^4$		    \\\hline
	\end{tabular}
\end{table}

\bibliography{bib/NewBib}
\end{document}